\documentclass[fleqn,useAMS,usenatbib]{mnras}

\usepackage{graphicx}% Including figure files
\usepackage{grffile}
\usepackage{amsmath}% Advanced maths commands
\usepackage{amssymb}% Extra maths symbols
\usepackage[dvipsnames]{xcolor}
\usepackage{color}
\usepackage{natbib}
\usepackage{lineno}

\usepackage[T1]{fontenc}
\usepackage{ae,aecompl}

\usepackage{times}
\usepackage{eso-pic}% http://ctan.org/pkg/eso-pic
\newcommand{\xivt}{\xi_{\rm vt}}
\newcommand{\xitt}{\xi_{\rm tt}}
\newcommand{\xitm}{\xi_{\rm tm}}
\newcommand{\xivg}{\xi_{\rm vg}}
\newcommand{\xivc}{\xi_{\rm vc}}
\newcommand{\xivm}{\xi_{\rm vm}}

\newcommand{\ximm}{\xi_{\rm mm}}
\newcommand{\rv}{r_{\rm v}}
\newcommand{\rvm}{\bar{r}_{\rm v}}
\newcommand{\bs}{b_{\rm slope}}
\newcommand{\co}{c_{\rm offset}}
\newcommand{\lrb}{b_{\rm rel}}
\newcommand{\magneticum}{\textsc{magneticum}}
\newcommand{\redmagic}{\textsc{redMaGiC}}
\newcommand{\redmapper}{\textsc{redMaPPer}}
\newcommand{\mice}{\textsc{mice}~2}
\newcommand{\vide}{\textsc{vide}}
\newcommand{\zobov}{\textsc{zobov}}

\title[Bias of void tracers in DES]{On the relative bias of void tracers in the Dark Energy Survey}

\AddToShipoutPictureBG*{%
  \AtPageUpperLeft{%
    \hspace{0.75\paperwidth}%
    \raisebox{-3.5\baselineskip}{%
      \makebox[0pt][l]{\textnormal{DES-2018-0340}}
}}}%

\AddToShipoutPictureBG*{%
  \AtPageUpperLeft{%
    \hspace{0.75\paperwidth}%
    \raisebox{-4.5\baselineskip}{%
      \makebox[0pt][l]{\textnormal{FERMILAB-PUB-18-220}}
}}}%

\author[DES Collaboration]{
\parbox{\textwidth}{
\Large
G.~Pollina$^{1,2}$\thanks{\href{mailto:gpollina@usm.lmu.de}{gpollina@usm.lmu.de}},
N.~Hamaus$^{2}$,
K.~Paech$^{1,2}$,
K.~Dolag$^{2,3}$,
J.~Weller$^{1,4,2}$,
C.~S{\'a}nchez$^{5,6}$,
E.~S.~Rykoff$^{7,8}$,
B.~Jain$^{5}$,
T.~M.~C.~Abbott$^{9}$,
S.~Allam$^{10}$,
S.~Avila$^{11}$,
R.~A.~Bernstein$^{12}$,
E.~Bertin$^{13,14}$,
D.~Brooks$^{15}$,
D.~L.~Burke$^{7,8}$,
A.~Carnero~Rosell$^{16,17}$,
M.~Carrasco~Kind$^{18,19}$,
J.~Carretero$^{6}$,
C.~E.~Cunha$^{7}$,
C.~B.~D'Andrea$^{5}$,
L.~N.~da Costa$^{16,17}$,
J.~De~Vicente$^{20}$,
D.~L.~DePoy$^{21}$,
S.~Desai$^{22}$,
H.~T.~Diehl$^{10}$,
P.~Doel$^{15}$,
A.~E.~Evrard$^{23,24}$,
B.~Flaugher$^{10}$,
P.~Fosalba$^{25,26}$,
J.~Frieman$^{10,27}$,
J.~Garc\'ia-Bellido$^{28}$,
D.~W.~Gerdes$^{23,24}$,
T.~Giannantonio$^{29,30,2}$,
D.~Gruen$^{7,8}$,
J.~Gschwend$^{16,17}$,
G.~Gutierrez$^{10}$,
W.~G.~Hartley$^{15,31}$,
D.~L.~Hollowood$^{32}$,
K.~Honscheid$^{33,34}$,
B.~Hoyle$^{4,2}$,
D.~J.~James$^{35}$,
T.~Jeltema$^{32}$,
K.~Kuehn$^{36}$,
N.~Kuropatkin$^{10}$,
M.~Lima$^{37,16}$,
M.~March$^{5}$,
J.~L.~Marshall$^{21}$,
P.~Melchior$^{38}$,
F.~Menanteau$^{18,19}$,
R.~Miquel$^{39,6}$,
A.~A.~Plazas$^{40}$,
A.~K.~Romer$^{41}$,
E.~Sanchez$^{20}$,
V.~Scarpine$^{10}$,
R.~Schindler$^{8}$,
M.~Schubnell$^{24}$,
I.~Sevilla-Noarbe$^{20}$,
M.~Smith$^{42}$,
M.~Soares-Santos$^{43}$,
F.~Sobreira$^{44,16}$,
E.~Suchyta$^{45}$,
G.~Tarle$^{24}$,
A.~R.~Walker$^{9}$,
W.~Wester$^{10}$
\begin{center} (DES Collaboration) \end{center}
}
\vspace{0.4cm}
\\
\parbox{\textwidth}{
Author affiliations are listed at the end of this paper}
}

\date{Accepted XXX. Received YYY; in original form ZZZ}

\pubyear{2018}

\begin{document}
\label{firstpage}
\pagerange{\pageref{firstpage}--\pageref{lastpage}}

\maketitle

\begin{abstract}
Luminous tracers of large-scale structure are not entirely representative of the distribution of mass in our Universe. As they arise from the highest peaks in the matter density field, the spatial distribution of luminous objects is biased towards those peaks. On large scales, where density fluctuations are mild, this bias simply amounts to a constant offset in the clustering amplitude of the tracer, known as linear bias. In this work we focus on the \emph{relative} bias between galaxies and galaxy clusters that are located inside and in the vicinity of cosmic voids, extended regions of relatively low density in the large-scale structure of the Universe. With the help of hydro-dynamical simulations we verify that the relation between galaxy and cluster overdensity around voids remains linear. Hence, the void-centric density profiles of different tracers can be linked by a single multiplicative constant. This amounts to the same value as the relative linear bias between tracers for the largest voids in the sample. For voids of small sizes, which typically arise in higher density regions, this constant has a higher value, possibly showing an environmental dependence similar to that observed for the linear bias itself. We confirm our findings by analysing mocks and data obtained during the first year of observations by the Dark Energy Survey. As a side product, we present the first catalogue of three-dimensional voids extracted from a photometric survey with a controlled photo-z uncertainty. Our results will be relevant in forthcoming analyses that attempt to use voids as cosmological probes.
\end{abstract}


%For comments:
%\noindent \GP{Giorgia}\\ \NH{Nico}\\ \KP{Kerstin}\\ \KD{Klaus}\\ \JW{Jochen}\\ \DES{Internal review}

%%%%%%%%%%%%%%%%%%%%%%%%%%%%%%%%%%%%%%%%%%%%%%%%%%

%%%%%%%%%%%%%%%%% BODY OF PAPER %%%%%%%%%%%%%%%%%%

%\begingroup
%\let\clearpage\relax
%\tableofcontents
%\endgroup
%\newpage

\section{Introduction}
\label{intro}

Most of the mass content in our Universe is composed of cold dark matter (CDM), currently described as a non-relativistic collisionless fluid which is responsible for the formation of halos, gravitationally bound clumps of dark matter that provide the potential wells in which baryons can cool and collapse to give birth to the galaxies we observe in the sky \citep{peebles1980}. While the quest for the nature of dark matter remains unresolved, currently the only way to infer its properties is indirect, via the gravitational interaction it exerts on the luminous constituents of the cosmos. To map the CDM one can therefore rely on the distribution of luminous tracers, such as galaxies and clusters of galaxies. Unfortunately, these objects are located only in the highest peaks of the underlying matter density field, therefore their clustering properties do not exactly mirror those of the CDM: galaxies and clusters of galaxies are biased tracers of the total mass distribution \citep{kaiser1984}. On small scales, where highly non-linear effects are important, this bias constitutes an unknown function of space and time. But on large scales, where density fluctuations remain within the linear regime, it can be modelled as a multiplicative offset in the clustering amplitude. The latter is known as {\it linear bias} and depends on a number of properties of the tracer population, one of the most important being the mass of its host halos: tracers residing in more massive halos exhibit a higher clustering bias \citep[for a comprehensive review, see][]{Desjacques2016}.

Typically, bias has been studied via the correlation function or the power spectrum of all tracers as a whole, regardless of their cosmic-web environment \citep[see, e.g.,][and references therein]{Smith:2007aa, Cacciato:2012aa, Springel:2018aa, Simon:2017aa, Dvornik:2018aa}. In a recent paper, however, \citet[][]{pollina2017} investigated the properties of bias focusing on tracers located in the vicinity of cosmic voids, large and relatively empty regions of large-scale structure. Voids, amongst all other structure types, are the largest in the Universe and make up the dominant fraction of its space. In \citet[][]{pollina2017} simulations were analysed to determine the absolute clustering bias of various tracers with respect to the total mass distribution. In order to mimic an observational approach, voids were identified in the distribution of tracers to define void catalogues. Then, both the density of dark matter particles and the density of the tracers themselves was investigated as a function the distance $r$ to the centers of these voids. In particular, it was found that the void-tracer cross-correlation function $\xivt(r)$ exhibits a linear relation with the corresponding void-matter cross-correlation function $\xivm(r)$, with a proportionality constant $\bs$,

\begin{equation}
\label{eq:voidBias}%
\xivt(r) = \bs \; \xivm(r).
\end{equation}
Furthermore, \citet{pollina2017} investigated the dependence of $\bs$ on void size. It was found that the best-fit value for $\bs$ decreases monotonically towards larger voids, and saturates to a constant number for the largest voids. This number was shown to coincide with the linear tracer bias $b_{\rm t}$, which can be either calculated from theory, or determined using the common bias estimators. Hence, $\bs$ in Equation~(\ref{eq:voidBias}) can be expressed as follows:

\begin{equation}
\bs(\rv)
\begin{cases}
> b_{\rm t}\;,\mathrm{for}\; \rv<\rv^+ \\
= b_{\rm t}\;,\mathrm{for}\; \rv\ge \rv^+\;,
\end{cases}
\end{equation}
where $\rv$ is the average, and $\rv^+$ the critical effective void radius of the sample. In other words, Equation~(\ref{eq:voidBias}) linearly relates tracer and matter densities around voids in all cases, but $\bs$ coincides with the linear bias $b_{\rm t}$ only when voids of size $\rv>\rv^+$ are considered in the measurement \citep[for visualization please refer to Fig. 4 of][]{pollina2017}. The precise value of $\rv^+$ depends on various properties of the tracer distribution itself, such as its sparsity and bias. Nevertheless, Equation~(\ref{eq:voidBias}) provides a very simple guideline of how to infer the distribution of mass around voids in the tracer distribution\footnote{Note that \citet{Nadathur2017} find a residual from the linearity of Equation~(\ref{eq:voidBias}) when $\bs$ is fixed to the linear bias $b_{\rm t}$, while \citet{pollina2017} and this paper treats it as a free parameter.}. The aim of this paper is to show that the same applies when relating different types of tracers around voids, both in simulations, and for the first time in observational data as well.

The importance of the result summarized in the previous paragraph can be better understood if we consider how building a coherent framework within void-cosmology has been so far a very complicated task. For example, obtaining accurate predictions on the most basic statistic, the void number counts, has always been particularly difficult. This has to do with the fact that the definition criteria for voids and their associated assumptions are not unique and typically differ between theory and practice. It is generally agreed that voids are vast regions of large-scale structure with a density below the average density of the Universe. However, due to the multi-scale nature of cosmic web, it is unclear how to divide local underdensities of different shape with multiple levels of nested substructure into a unique set of distinct objects. In a pioneering theoretical study, \citet{sheth2004} define voids as spherically symmetric underdensities that undergo shell crossing at their boundaries. Their initial density profile is assumed to have an inverted top-hat form, and spherical evolution is adopted to predict the final void abundance following the excursion-set formalism~\citep[also see][]{jennings2013,Nadathur2015,falck2015,Chan:2014aa}. In practice, however, these assumptions are hardly ever justified.
Two general directions have been pursued to overcome this problem. One is to modify or relax specific assumptions in the theory of \citet{sheth2004}, such as demanding volume conservation for the entire void sample~\citep{jennings2013}, or allowing the critical density threshold for void formation to vary as a free parameter~\citep{pisani2015}. The other option is a modification of void catalogues via selection cuts, which guarantee the assumptions in \citet{sheth2004} to be satisfied~\citep{ronconi2017}. Both approaches show promising results and will likely play a role in future analyses that attempt to extract cosmological signals from voids. 

However, theoretical calculations rely on a smooth matter-density field to define voids, while observations can only provide a discrete distribution of tracers in three dimensions. 
A number of different methods have recently been developed to quantitatively extract void catalogues from observations~\citep[see, e.g.,][]{Padilla_etal_2005, neyrinck2008, VIDE2015}, but their full connection to theory remains an open problem. Given the large number of observational void catalogues already published~\citep[][]{pan2012, sutter2012SDSSdr7, ceccarelli2013, mao2016, nadathur2016BOSS}, and expected to become available with future surveys~\citep[e.g., LSST, EUCLID, DESI, see][respectively]{LSST,EUCLID,DESI}, it is important to address this issue. 
The results of \citet{pollina2017} provide a first step to connect theory with practice, as Equation~(\ref{eq:voidBias}) allows us to bridge the gap between the matter- and tracer density profiles around observationally defined voids. In fact, these results have already been employed to this end by \citet{ronconi2017}, who extended their theoretical void size function to voids traced in halos thanks to Equation (\ref{eq:voidBias}). 

While the first models for void evolution \citep[][]{hausman1983,bertschinger1985} have been developed soon after their earliest observations \citep[][]{GregoryThompson1978, kirshner1981}, the previous decade has witnessed an increasing number of publications unveiling the potential of various void properties to provide new insights into cosmology. For example, their average density profile has been shown to follow a universal shape across void size, redshift, and tracer type that can be described by a narrow family of empirical functions \citep[e.g.,][]{ricciardelli2013, ricciardelli2014, hamaus2014, sutter2014sparseS}. Based on the cosmological principle, voids represent a population of statistically isotropic spheres distributed at different redshifts, allowing us to probe the expansion history of the Universe by means of the Alcock-Paczynski (AP) test \citep{AlcockPaczynski1979,lavaux2012, sutter2012, sutter2014, hamaus2014b, hamaus2016, mao2016}. It has been investigated whether the observed Cold Spot in the Cosmic Microwave Background (CMB) could be explained as Integrated Sachs-Wolfe (ISW) imprint caused by very large voids along the line of sight \citep[e.g.,][]{rees1968, Finelli_etal_2014, kovac2014, Nadathur2015coldspot, Kovacs:2018aa}, and a final conclusion on this topic is yet to be reached; the potential of the ISW by voids is nevertheless important and still being actively investigated \citep[e.g.][]{Granett:2008aa,Cai:2014aa,Nadathur:2016aa,Kovacs:2017aa}. It has further been argued that void number counts have the potential to improve on dark energy constraints \citep[][]{pisani2015} and that together with their average density profile can discriminate modify gravity \citep[][]{Li_2011, clampitt2013, cai2016, Barreira_etal_2015, zivick2015}, coupled dark energy \citep[][]{pollina2015}, and massive neutrino \citep[][]{massara2015} cosmologies from $\Lambda$CDM. These probes could also be sensitive to possible degeneracies between warm dark matter and modifications of gravity \citep{baldi2016}. Most recently, redshift-space distortions (RSD) around voids have been identified as a promising source of additional cosmological information~\citep{hamaus2015, hamaus2016, cai2016, chuang2016, achitouv2016, Hawken:2017aa, achitouv2017, hamaus2017}. In order to fully exploit the associated signal, a reliable model for tracer bias in void environments is indispensable, which is the subject of \citet{pollina2017} and this work.

Despite the fact that Equation~(\ref{eq:voidBias}) has a number of interesting consequences and applications, it is challenging to test experimentally, as the dark matter density cannot be observed directly in all three dimensions. However, voids can also be used as weak gravitational (anti-)lenses to infer their projected surface mass density \citep{krause2013, Melchior:2014aa, clampitt2014, sanchezDES2016}. Either a deprojection of the void lensing profiles to 3D, or a projection of tracer density profiles to 2D then allows us to constrain the bias relation in voids (Fang et al., in prep). Another possibility is to apply Equation~(\ref{eq:voidBias}) to different tracers of the matter distribution. As long as every individual tracer obeys a linear clustering bias with respect to the dark matter, the relative clustering bias between the tracers should remain linear as well. In this analysis we will make use of galaxies and galaxy clusters as two distinct tracer types. These are the most commonly available and abundant tracers in current surveys, and at the same time exhibit very different clustering properties. We will use the distribution of galaxy \emph{clusters} to define our void sample, thanks to their higher fidelity in providing photometric redshifts and thus accurate distance estimates. The relative bias relation between galaxies and galaxy clusters will be thoroughly investigated in the vicinity of those voids. In order to provide a controlled setup, we first develop our analysis techniques based on state-of-the-art hydrodynamical simulations (\textsc{magneticum}). Our methods are then applied to the \redmagic\ galaxy-- and \redmapper\ galaxy cluster catalogues originating from the first year of observations by the DES collaboration. Realistic mock catalogues provided by the \mice\ project that have been constructed to specifically mimic the observations which will be used to validate our conclusions.

This paper is organized as follows: in section \ref{data} we present all the data employed in our study (hydro-sims, DES mocks and DES data); in section \ref{methods} we describe the void finding algorithm, as well as all the methods employed to estimate the relative bias of tracers; in section \ref{analysis} we present the results of our analysis; finally we discuss our conclusions in section \ref{Concl}.

\section{Simulations, data and mocks}
\label{data}

\subsection{Simulations}
\label{sims}

\begin{table}
	\centering
	\caption{Properties of the galaxy and cluster samples in the \magneticum\ simulations. The minimum mass $M_{\rm min}$ is given in terms of stellar mass $M_{*}$ for galaxies, and in terms of $M_{\rm 500c}$ for clusters. $N_{\rm t}$ is the total number of tracers and $N_{\rm v}$ the corresponding number of identified voids.}
	\label{tab:tracers}
	\begin{tabular}{l|lcr} % 3 columns, alignment for each
		Tracers  & $M_{\rm min} [M_{\sun}/h]$    & $N_{\rm t}$          &  $N_{\rm v}$        \\
		\hline\hline
                Galaxies &  $M_{*} =  \, 1 \times 10^{11}$    				& $6.5 \times \, 10^6$ &  -      \\
                 				&  $M_{*}=  5 \times 10^{11}$         & $2.6 \times \, 10^6$ &  -      \\  
                         		&  $M_{*}= 1 \times 10^{12}$ 						& $3.5 \times \, 10^5$ &  -       \\
               \hline
               Clusters   	&  $M_{\rm 500c}= 1 \times 10^{14}$         			& $1.0 \times \, 10^5$ &  $1053$       \\
		\hline
	\end{tabular}
\end{table}

The hydrodynamical simulation suite \magneticum\ {\it pathfinder}\footnote{\url{http://www.magneticum.org}} (Dolag et al, in prep.) has already been employed successfully in a wide number of numerical studies. \magneticum\ showed so far a remarkably good agreement with observations for various probes, such as for the pressure profiles of the intra-cluster medium \citep{2013A&A...550A.131P,2014ApJ...794...67M}, the expected Sunyaev Zeldovich signal \citep{dolag2015}, the imprint of the intergalactic medium onto the dispersion signal of Fast Radio Bursts \citep{dolag2015FastRB}, various characteristics of AGN populations \citep{hirschmann2014,2015MNRAS.448.1504S,2016MNRAS.458.1013S}, the dynamical features of massive spheroidal galaxies \citep{2013ApJ...766...71R,2016arXiv160506511R}, and the angular momentum signatures of galaxies \citep{teklu2015}. 

In this work we employ the largest cosmological volume simulated within that project, it covers a box of side length $2688h^{-1}$~Mpc, simulated using $2\times4536^3$ particles \citep[for details, see][]{2016MNRAS.456.2361B}. We adopted a WMAP7 \citep{Komatsu11} $\Lambda$CDM cosmology with $\sigma_8 = 0.809$, $h = 0.704$, $\Omega_\Lambda = 0.728$, $\Omega_m = 0.272$, $\Omega_b = 0.0456$, and an initial slope for the power spectrum of $n_s = 0.963$.
The simulation is based on {\small P-GADGET3}~\citep{Springel:2005aa}, a parallel cosmological tree Particle-Mesh (PM) Smoothed-Particle Hydrodynamics (SPH) code. It uses an entropy-conserving formulation of SPH \citep{2002MNRAS.333..649S} and follows the gas using a low-viscosity SPH scheme to properly track turbulence \citep{2005MNRAS.364..753D}.    	
Halos and sub-halos are identified using the {\sc subfind} algorithm \citep{2001MNRAS.328..726S,2009MNRAS.399..497D}. {\sc subfind} identifies sub-structures as locally overdense, gravitationally bound groups of particles, starting from a main halo which is identified through the Friends-of-Friends (FoF) algorithm with a linking length of 0.16 times the mean inter-particle separation. After this first step a local density is estimated for each particle via adaptive kernel estimation, making use of a prescribed number of smoothing neighbours. After isolated density peaks are identified, additional particles of decreasing density are added. When a saddle point that connects two disjoint overdense regions in the global density field is reached, the smaller structure between the two is treated as a sub-structure candidate, and the two overdensities are then merged. An iterative unbinding procedure with a tree-based calculation of the potential is then run on all sub-structure candidates. These structures are finally associated with galaxies, and their integrated properties (such as stellar mass, $M_*$) are computed. Galaxies in \magneticum\ can have stellar masses as low as $4 \times 10^8h^{-1} M_{\sun}$, but in this study we will consider as main sample only galaxies with $M_*\ge10^{11}h^{-1} M_{\sun}$, which are more realistically observable.
 
The virial radius of the main haloes identified by the FoF algorithm is calculated using a density contrast built on the top-hat model \citep{Eke96}. To allow a better comparison with observations, we additionally use an overdensity with respect to 500 times the critical density to define $M_{\rm 500c}$, which is the mass we will refer to as cluster mass in this paper. Clusters are identified as main haloes with $M_{\rm 500c} > 10^{13}h^{-1} M_{\sun}$, but in this paper we only consider \magneticum\ clusters above $10^{14}h^{-1} M_{\sun}$. For our analysis we make use of the galaxy and cluster samples extracted from the simulation at redshift $z=0.14$ with the criteria explained above. In Table \ref{tab:tracers} we summarise some properties of the tracers relevant in this work.

\subsection{Data}
\label{des}
The Dark Energy Survey \citep[DES, see][]{DES2005} is an on-going 5 year observational campaign supported by an international collaborative effort. It employs the $570$ megapixel Dark Energy Camera \citep[DECam, see][]{DECam2008, DECam2015} mounted on the Blanco telescope at the Cerro Tololo Inter-American Observatory (CTIO). At the end of its operations, DES will have mapped approximately $300$ million galaxies and tens of thousands of clusters over a 5000 square degree footprint in the southern hemisphere. DES provides photometric data using five filters ({\it grizY}) to the limiting magnitude of $24$th {\it i}-band \citep{DES2015imagingpipeline}, although the relevant limiting magnitude for this study is $22.5$ in {\it i}-band, as it constrains the observations of galaxies \citep{Drlica-Wagner:2018aa}. In this work we employ data obtained during the first year of observation (Y1) taken between Aug. 31 2013 and Feb. 9 2014, that have already shown their potential in constraining cosmology \citep{DES2017ccfgcwl}. DES Y1 wide-field observations scanned a large region extending approximately between $-60^{\circ} < \delta <-40^{\circ}$ overlapping the South Pole Telescope (SPT) survey footprint, screening an area of $1321 \, \rm{deg}^2$ (A1). A much smaller area overlapping the ``Stripe 82" of the Sloan Digital Sky Survey (SDSS) was also mapped by DES, but this region will not be included in our analysis. From the Gold catalogues \citep{Drlica-Wagner:2018aa}, $26$ million galaxies were selected for the weak lensing sample.   
Recently the first three years of the observational campaign were made public with the first DES data release \citep{DESDR1}.

\subsubsection{Galaxy clusters}
We make use of \emph{red-sequence Matched-filter Probabilistic Percolation} (\redmapper) Y1A1 clusters \citep{McClinton-Vargas:2018a}, both to use them as tracers of the large-scale structure, and to identify cosmic voids in the latter. The photometric red-sequence cluster finder \redmapper\ is specifically developed for large photometric surveys. It identifies galaxy clusters by searching for a bulk of its population to be made up of old, red galaxies with a prominent 4000\AA-break. Focusing on this specific galaxy population the algorithm increases the contrast between cluster and background galaxies in colour space, and it enables accurate and precise photometric redshift estimates, with a scatter of $\sigma_z /(1+z) = 0.01$ level for $z < 0.7$ \citep{redmapperSV}, which includes the redshift window employed for data analysis in this paper. The associated cluster richness estimator, $\lambda$, is the sum of the membership probability of every galaxy in the cluster field, and has been optimized to reduce the scatter in the richness-mass relation \citep{Rozo2009, Rozo2011, Rykoff2012a}. For a more detailed description of the algorithm we refer to \cite{redmapperSV}. In this work we will employ cluster samples with $\lambda > 5$, which corresponds to a minimum mean mass of about $\sim10^{13}h^{-1} M_{\sun}$ following the mass-richness relation of \citet{McClinton-Vargas:2018a}. This low richness cut that does not guarantee the purest cluster selection. In this paper, however, we are not interested in the detailed properties of individual clusters. Rather, we desire the selected sample to be used as a tracer of large-scale structure, regardless of whether some of its objects are true clusters or not. The resulting full catalogue contains $103423$ clusters and has proven to be optimal for the task of void identification, owing to its relatively high cluster density of about $10^{-4} h^3\rm{Mpc}^{-3}$.

\subsubsection{Galaxies}
We also employ \emph{red-sequence Matched-filter Galaxy Catalog} (\redmagic) Y1A1 galaxies \citep{Elvin-Poole:2017aa} as tracers of large-scale structure. The \redmagic\ algorithm \citep{redmagicSV} is automated for selecting Luminous Red Galaxies (LRGs) and was specifically designed to minimize photometric redshift uncertainties in photometric large-scale structure studies, resulting in a photo-z bias $z_{\rm spec} - z_{\rm photo}$ better than $0.005$ and in a scatter $\sigma_z/(1+z)$  of $0.017$. \redmagic\ achieves this goal by self-training the colour cuts necessary to produce a luminosity-thresholded LRG sample of constant comoving density. In this work we will distinguish among three different \redmagic\ samples, denoted as \emph{high density} (brighter than $0.5 \, L_*$ and density  $10^{-3} h^3\rm{Mpc}^{-3}$), \emph{high luminosity} (brighter than $1 \, L_*$ and density $4 \times 10^{-4} h^3\rm{Mpc}^{-3}$), and \emph{higher luminosity} (brighter than $1.5 \, L_{*}$ and density $10^{-4} h^3\rm{Mpc}^{-3}$).

\subsection{DES Mocks}
\label{mocks}
In order to validate our results, we make use of mock catalogues extracted from the \mice\ project. \mice, based on the original \textsc{mice} (MareNostrum - Instituto de Ciencias del Espacio) project \citep{MICE2015, Fosalba2015}, is a suite of large high-resolution $N$-body simulations that have been run with the \textsc{gadget 2}\ code \citep{Springel:2005aa}. Including $4096^3$ particles in a box size of $3.072h^{-1}$Gpc, \mice\ resolves halos with even lower mass resolution ($2.93 \, \times 10^{10} \, h^{-1} M_{\sun}$) than \textsc{mice}, making this particular simulation a perfect tool in providing mocks for deep and sensitive surveys such as DES. FoF halo catalogues extracted from the simulations are populated by galaxies using a Halo Occupation Distribution (HOD), which assigns luminosities to the central and satellite galaxies so that their observed luminosity function is preserved. The \mice\ galaxy catalogue is forced to match luminosity, colours and clustering properties of DES at redshift $z=0.1$, from where a light cone is then extrapolated by replicating and translating the simulation box, allowing one to build an output with negligible repetition up to redshift $z=1.4$. In this work we are going to employ the largest available light cone, which reproduces a full octant of the sky with the same properties as the DES Y1 observations, such as photometry. More specifically, we will employ the \redmagic\ galaxy and \redmapper\ cluster catalogues extracted from \mice\ to asses the impact of photometric redshift uncertainty on our results.

\section{Methods}
\label{methods}
%In this section we present the void finder and the statistical tools that are employed in this paper.

\subsection{Void finder}
\label{finder}
We employ the Void IDentification and Examination toolkit \vide\ \citep{VIDE2015} to construct our void catalogues. \vide\ implements an enhanced version of \zobov\ \citep[ZOnes Bordering On Voidness, ][]{neyrinck2008}, an algorithm that identifies density depressions in a $3$-dimensional set of points. The void finding procedure consists of three steps. First, the finder reads in the tracer positions and associates to each tracer a cell of volume that is closer to it than to any other tracer. This procedure is unique and referred to as \emph{Voronoi tessellation}, the resulting cells are denoted \emph{Voronoi cells}. By assuming equal weights for all particles it is straight-forward to associate a density to each Voronoi cell: it is simply obtained as the inverse of the Voronoi cell volume. In this manner every point inside the tracer distribution can be associated with a density, hence a well-defined density field is obtained. As a second step, local density minima are found and their surrounding basins identified. A local density minimum is a Voronoi cell of given volume whose neighbouring cells all have smaller volumes, respectively higher densities, than the central cell. Starting from these density minima, surrounding Voronoi cells are merged consecutively if their individual density is above the one of the previously merged cell. Once a cell of lower density is encountered, the process of merging is stopped. Thus, this procedure delineates local density basins, denoted as \emph{zones}, with their surrounding ridges in the tracer distribution.

Finally, zones are merged to become voids by means of the so-called \emph{watershed} algorithm \citep[e.g.,][]{Platen2007}. To this end a density threshold is raised starting from each zone's local density minimum. In analogy to a rising water level on a two-dimensional terrain, water flows into adjacent zones when the separating ridges are overflown. As long as shallower zones are added to the original zone, the final void consists of all such merged zones, which are still recorded as its sub-voids. When a deeper zone is encountered, the process is stopped. Therefore, the watershed algorithm naturally constructs a hierarchical structure of nested voids. Optionally, in order to prevent including very overdense structures inside voids, a density threshold for ridge densities can be set. It is typically chosen to be $20\%$ of the mean tracer density.

In this work we will employ the most general void catalogue produced by \vide, without applying any further selection cuts on density or hierarchy levels of voids. We define the void centre as the volume-weighted barycentre $\vec{X}$ of the $N$ Voronoi cells that define each void,

\begin{equation}
  \label{eq:v-centre}
  \vec{X} = \left.\sum\limits_{i=1}^N \vec{x}_i \cdot V_i \right/ \sum\limits_{i=1}^N V_i  \;,
\end{equation}
where $\vec{x}_i$ are the coordinates of the $i$-th tracer of that void, and $V_i$ the volumes of their associated Voronoi cells. The \emph{effective} void radius $\rv$ is calculated from the total volume of the void $V_{\rm v}$. It is defined as the radius of a sphere with the same volume,

\begin{equation}
\label{eq:v-vol}
  V_{\rm v} \equiv {\sum\limits_{i=1}^N V_i } = \frac{4\pi}{3}\rv^3 \,.
\end{equation}

\subsection{Correlation functions}
\label{densProf}
In order to explore the clustering statistics around voids we will employ correlation functions. The two-point correlation function $\xi_\mathrm{t_1t_2}(r)$ between a tracer t$_1$ and a tracer t$_2$ is defined via the ensemble average

\begin{equation}
\xi_\mathrm{t_1t_2}(r) \equiv \langle \delta_{\rm t_1}(\vec{x})\delta_{\rm t_2}(\vec{x}+\vec{r}) \rangle\;,
\end{equation}
where the spatial density fluctuation of a tracer around its average density $\langle n_{\rm t} \rangle$ is given by $\delta_{\rm t}(\vec{x})=n_{\rm t}(\vec{x})/\langle n_{\rm t} \rangle - 1$.
%The correlation function $\xi_\mathrm{AB}(r)$ between a tracer A and a tracer B, distributed within volume elements $\delta V_\mathrm{A}$ and $\delta V_\mathrm{B}$, is defined via the excess probability above random to find a pair of A and B at separation $r$,
%\begin{equation}
%\delta P = \langle n_{\rm A} \rangle \langle n_{\rm B} \rangle\left[1+\xi_\mathrm{AB}(r)\right]\delta V_\mathrm{A}\delta V_\mathrm{B}\;,
%\end{equation}
%where $\langle n_{\rm A} \rangle$ and $\langle n_{\rm B} \rangle$ are the mean tracer densities at a given redshift~\citep{peebles1980}.
In the case where ${\rm t_1}={\rm t_2}$, this statistic is referred to as \emph{auto}-correlation function, and for ${\rm t_1}\neq{\rm t_2}$ as \emph{cross}-correlation function. The cross-correlation function between void centres and tracers $\xivt(r)$ is of particular relevance for this work. It can be shown to be equivalent to the average (or stacked) tracer-density profile of voids, $n_{\rm vt}(r)/\langle n_{\rm t} \rangle -1$~\citep[see][]{hamaus2015}. In simulations that incorporate periodic boundary conditions it is straight-forward to calculate; one simply histograms the number of tracers in spherical shells of width $\delta r$ around each void centre,

\begin{equation}
\label{eq:prof-xi}
n_{\rm vt}(r) =  \sum_i \frac{\Theta(|\vec{X} -\vec{x}_i| - r)}{\delta V(r)}\;,
\end{equation}
and then averages it over all voids. Here $\Theta$ represents a step function with

\begin{equation}
\Theta(x) = 
\begin{cases}
1\;,\mathrm{for}\,\, -\delta r/2 < x < \delta r/2 \\
0\;,\mathrm{otherwise}\;.
\end{cases}
\end{equation}
In order to suppress discreteness noise, we choose to keep the radial shell bins fixed in units of the void radius $\rv$ of each void, and normalize by the constant shell volumes

\begin{equation}
 \delta V(r) = \frac{4\pi}{3}\left[(r+\delta r/2)^3-(r-\delta r/2)^3\right](\rvm/\rv)^3
\end{equation}
after averaging over the tracer counts around all voids. The mean effective radius $\rvm$ of the void sample is used to rescale from dimensionless to physical units of volume.

However, in real observations we are observing tracers inside irregular boundaries of a survey mask on the past light cone. In that situation it is helpful to employ a catalog of randoms to isolate true from fake correlations in the data. To this end the \emph{Landy-Szalay} estimator~\citep{landy1993} provides a way to calculate the void-tracer cross-correlation function from data catalogues $D$ and random catalogues $R$ for each tracer and void sample,

\begin{equation}
\label{eq:LS}
 \xivt(r) = \frac{\langle D_{\rm v} D_{\rm t} \rangle - \langle D_{\rm v} R_{\rm t} \rangle - \langle D_{\rm t} R_{\rm v} \rangle + \langle R_{\rm v} R_{\rm t} \rangle}{\langle R_{\rm v} R_{\rm t} \rangle} \;,
\end{equation}
where angled brackets symbolize normalized pair counts at separation $r$ in units of $\rv$. They can be calculated as histograms in analogy to Equation~(\ref{eq:prof-xi}).

Void density profiles have been studied in detail in the recent literature~\citep[e.g.,][]{colberg2005, ricciardelli2013, ricciardelli2014, hamaus2014, sutter2014sparseS}. They typically exhibit a few very characteristic features: a deep under-dense core in the very centre, and an over-dense ridge (compensation wall) close to the effective radius $\rv$. The following empirical function was shown to capture these features accurately~\citep{hamaus2014},

\begin{equation}
  \label{eq:prof}
  \frac{n_{\rm vt}(r)}{\langle n_{\rm t} \rangle} -1 = \delta_c \frac{1-(r/r_s)^\alpha}{1+(r/\rv)^\beta}\;,
\end{equation}
where $\delta_c$ is the central density contrast at $r=0$, $r_s$ a scale radius at which the density equals the average density of tracers $\langle n_{\rm t} \rangle$, and $\alpha$, $\beta$ describe the inner and outer slopes of the profile. 

\subsection{Bias estimation}
\label{bias_estimate}

%\subsubsection{Hydro-simulations}
In simulations the clustering bias of any tracer can directly be calculated, because the dark matter particle locations are available. Therefore, it is simply given by the ratio of tracer and matter correlation functions,

\begin{equation}
\label{eq:galbias}
b_{\rm t} = \sqrt{\frac{\xitt(r)}{\ximm(r)}} \simeq \frac{\xitm(r)}{\ximm(r)}\;.
\end{equation}
The second equality only holds on large scales in the linear regime, where $b_{\rm t}$ is a constant number. In a similar manner we can define the relative bias between a tracer t$_1$ and a tracer t$_2$ as

\begin{equation}
\label{eq:relbias}
b_{\rm rel} \equiv \frac{b_{\rm t_1}}{b_{\rm t_2}} = \sqrt{\frac{\xi_{\rm t_1t_1}(r)}{\xi_{\rm t_2t_2}(r)}} \simeq \frac{\xi_{\rm t_1t_2}(r)}{\xi_{\rm t_2t_2}(r)}\;,
\end{equation}
where, without loss of generality, we may choose tracer t$_1$ to be the more highly biased one, such that $b_{\rm rel}>1$. In this analysis we will associate the highly biased tracer with galaxy clusters, and the less biased tracer with galaxies. %In particular, we will compare the values of $\bs$ obtained via Equation~(\ref{eq:voidBias}) with the linear bias as determined from Eqs.~\ref{eq:galbias} and \ref{eq:relbias}.

%\subsubsection{DES mocks and data}
In observational data, where we do not have direct access to the mass distribution, the absolute clustering bias of tracers can only be determined indirectly. We follow the approach of \citet{Paech:2017aa} and calculate the angular power spectra between tracer t$_1$ and tracer t$_2$ using the public code \textsc{class}\footnote{\url{http://class-code.net}}~\citep{Blas:2011aa} and its extension \textsc{class}gal~\citep{Di-Dio:2013aa},

\begin{equation}
\label{eq:class_cl}
C_\ell^{\rm t_1t_2} = 4\pi \int\frac{{\rm d} k}{k} P_{\rm ini}(k) \Delta_\ell^{\rm t_1}(k)\Delta_\ell^{\rm t_2}(k) \;.
\end{equation}
Here, $P_{\rm ini}(k)$ is the dimensionless primordial power spectrum at wavenumber $k$ and

\begin{equation}
\Delta^{\rm t}_\ell(k) = \int {\rm d} z \: b_{\rm t} \frac{ {\rm d} N_{\rm t}(z)}{{\rm d} z} j_\ell\left[k\, r(z)\right] D(z)T(k) \;,
\end{equation}
where ${\rm d} N_{\rm t}(z)/{\rm d} z$ is the redshift distribution and $r(z)$ the comoving distance of tracer t, $j_\ell$ the spherical Bessel function, $D(z)$ the growth factor, and $T(k)$ the transfer function. Assuming a fiducial flat $\Lambda$CDM cosmology with the parameters $h=0.678$, $\Omega_{\rm b}=0.048$, $\Omega_{\rm m}=0.308$, $\sigma_8=0.826$, $z_{\rm re}=11.3$ and $n_{\rm s}=0.96$~\citep{Planck-Collaboration:2014aa}, we can then infer the effective values of $b_{\rm t_1}$, $b_{\rm t_2}$ and their ratio (averaged within the considered redshift range) from the angular auto-power spectra of the two tracers. The angular power spectra are determined using the public code \textsc{polspice}\footnote{\url{http://www2.iap.fr/users/hivon/software/PolSpice}}~\citep{Szapudi:2001aa, Chon:2004aa} from a pixellated map of the projected tracer-density contrast on the sky. As in \citet{Paech:2017aa}, we treat the shot noise contribution to the angular power spectra as a free parameter, and consider a multipole range of $20<\ell<500$. The covariance of the $C_\ell$'s is estimated via applying a jack-knife sampling of the map, splitting up the map area into 100 contiguous regions of equal size.

\section{Analysis}
\label{analysis}
In this section we present the results of our analysis pipeline, applied to \magneticum\ simulations, \mice\ mocks, and finally DES data. We emphasize that all void catalogues employed in this paper are identified in the cluster samples at hand, regardless of the nature of the data set analysed. If needed, we refer to those voids as \emph{cluster-voids}, to distinguish them from voids identified in a different tracer population\footnote{The procedure can also be inverted, i.e. it is possible define voids in the galaxy sample and then use those voids to measure the density of galaxies and clusters around them. For consistency with the approach in \citet{pollina2017}, and for the advantage that will be presented in section \ref{photoz-unc}, we use the more highly biased tracer to identify voids.}.

\subsection{Hydro-simulations}
\label{analysis:sims}

\begin{figure*}
	\includegraphics[width=0.47\textwidth]{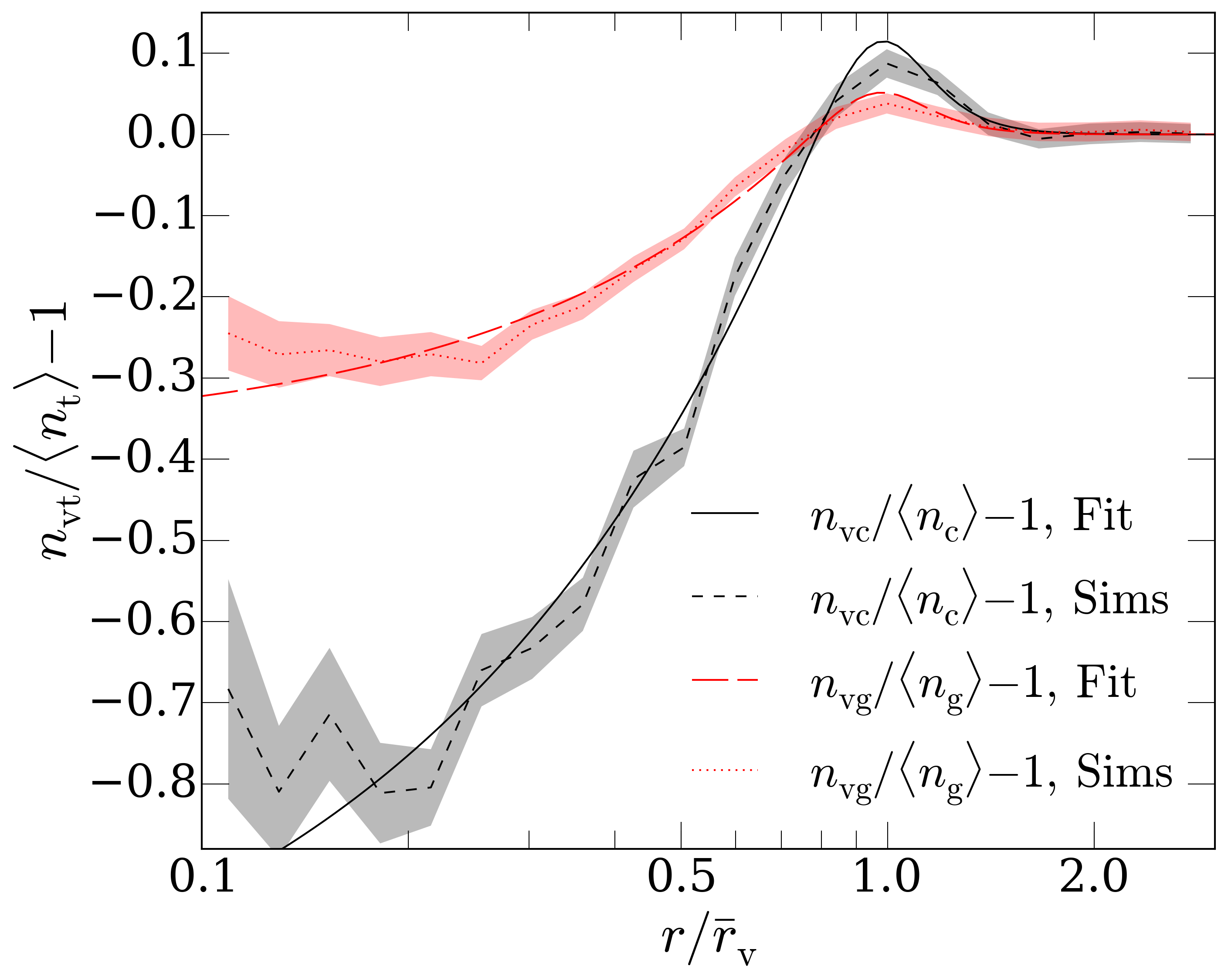}
	\includegraphics[width=0.47\textwidth]{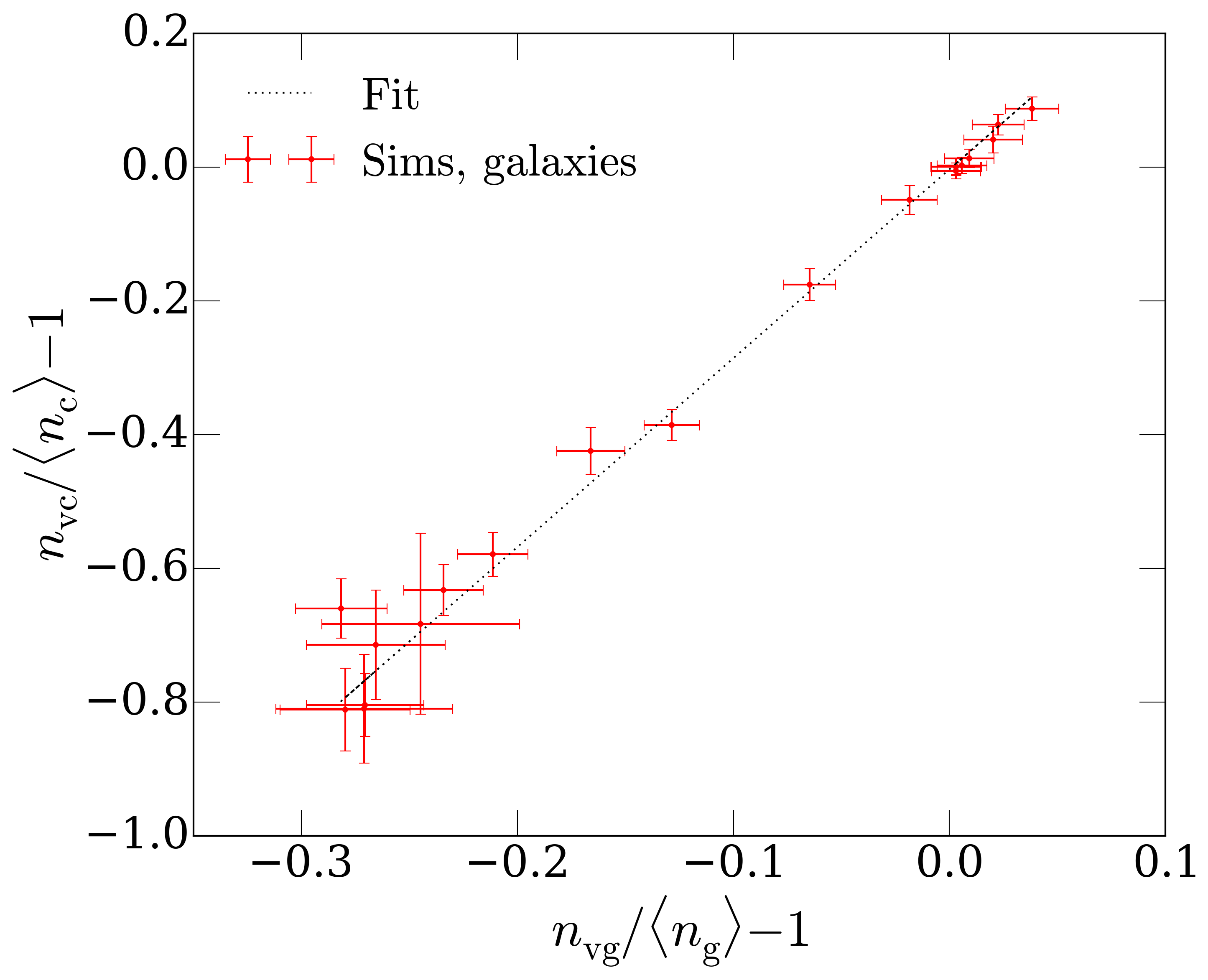}
	\caption{Left: Tracer-density profiles (dashed black for clusters, dotted red for galaxies) around cluster-defined voids of radius $190h^{-1}$Mpc $<\rv<220h^{-1}$Mpc in the \magneticum\ simulation. Solid black and long-dashed red lines show the best fits obtained via Equation~(\ref{eq:prof}). Right: Cluster- and galaxy-density profiles from the left panel plotted against each other (black points with error bars). The dotted black line shows the best fit using Equation~(\ref{eq:lin-fit}).}
	\label{fig:magn-prof}
\end{figure*}

With the help of the fully hydro-dynamical simulations \magneticum\ we investigate whether it is possible to use a similar approach to that presented by \citet{pollina2017}, albeit only considering clusters and galaxies as tracers. In this manner the relative bias is expected to obey similar properties as the linear bias, in analogy to Equation~(\ref{eq:voidBias}). The idea is to only use void catalogues that are defined in the most highly biased population available and then compute the average tracer-density profiles around voids of similar size using clusters and galaxies separately. The latter are hence exclusively used to compute galaxy-density profiles around cluster-voids.

We apply a conservatively high mass cut of $M_{\rm min} = 10^{14}h^{-1} M_{\odot}$ to our \magneticum\ clusters, firstly to make sure that we do not include objects that are of too low detection significance in the observed data, and secondly to achieve a relative bias between our cluster and galaxy sample that is significantly larger than unity. Since the lower limit for the bias of the galaxy sample is set by the mass resolution of the simulation, we can only boost the relative bias by increasing $M_{\rm min}$ for the cluster sample. This implies a lower resolution for smaller voids due to tracer sparsity~\citep[for further details on sparse sampling and void finding, see][]{sutter2014sparseS}, so the resulting void catalogue contains rather large objects. However, as we are only interested in the relation between tracer-density profiles around a fixed void population, the absolute distribution of void sizes does not matter for our purposes.

In the left panel of Fig.~\ref{fig:magn-prof} we show the stacked density profile of such cluster-voids computed twice: once using the same cluster population they were identified in (dashed black line), and once using the full galaxy sample extracted from \magneticum\ (red dotted line). The shaded areas represent the uncertainty on the mean density profile, computed as the standard deviation of all individual void profiles from their mean. The void density profiles are calculated following the procedure explained in the beginning of Section~\ref{densProf}, including voids of effective radii in the range $190h^{-1}$Mpc $<\rv<220h^{-1}$Mpc. The function from Equation~(\ref{eq:prof}) is used to fit the density profiles (solid black for clusters and long-dashed red for galaxies), yielding a good match in both cases. This corroborates the universal character of Equation~(\ref{eq:prof}) with respect to tracer type. The very characteristic features are a clear under-dense core close in the void centre and a compensation wall around $r\simeq\rv$, which are most pronounced in the cluster-density profile. When the density profile of galaxies around the same cluster-voids is computed, those features are less pronounced, but still clearly visible. Because clusters have a higher clustering bias than galaxies, this behaviour is expected.

\begin{figure*}
	\includegraphics[width=0.99\textwidth]{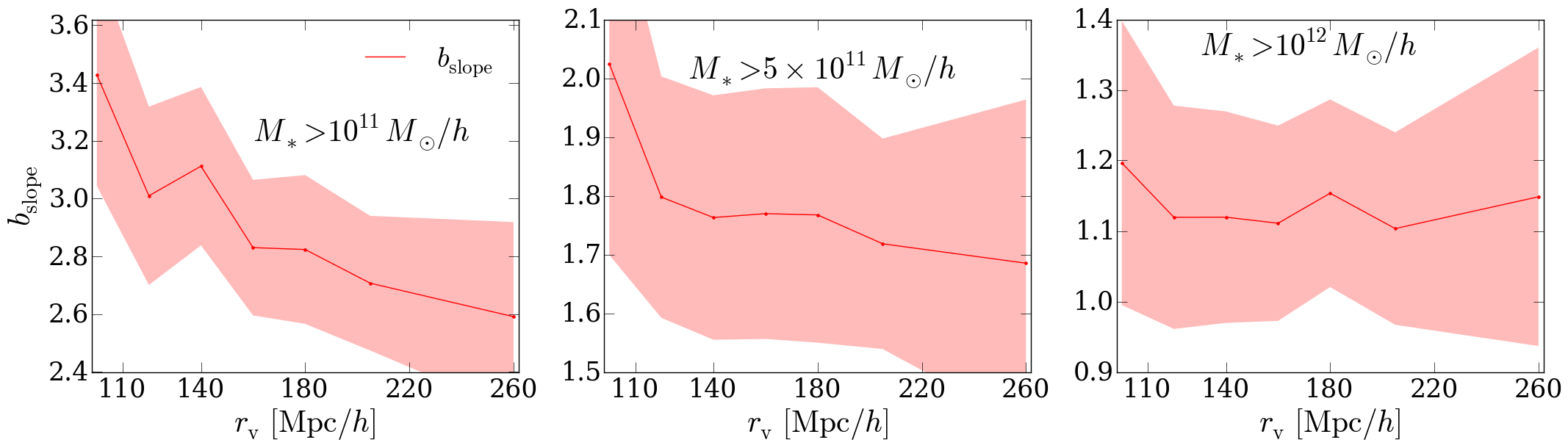}
	\caption{Best-fit values for $\bs$ as a function of effective void radius in the \magneticum\ simulation. The stellar-mass cut for the galaxy sample is varied from left to right, as indicated in each panel. The cluster sample has a fixed mass cut of $M_{\rm 500c} > 10^{14}h^{-1} M_{\odot}$, it is also used for the void identification.}
	\label{fig:magn-bias-vs-size}
\end{figure*}

\begin{figure*}
	\includegraphics[trim=0. 0. 0.2cm 0., clip, height=5.2cm]{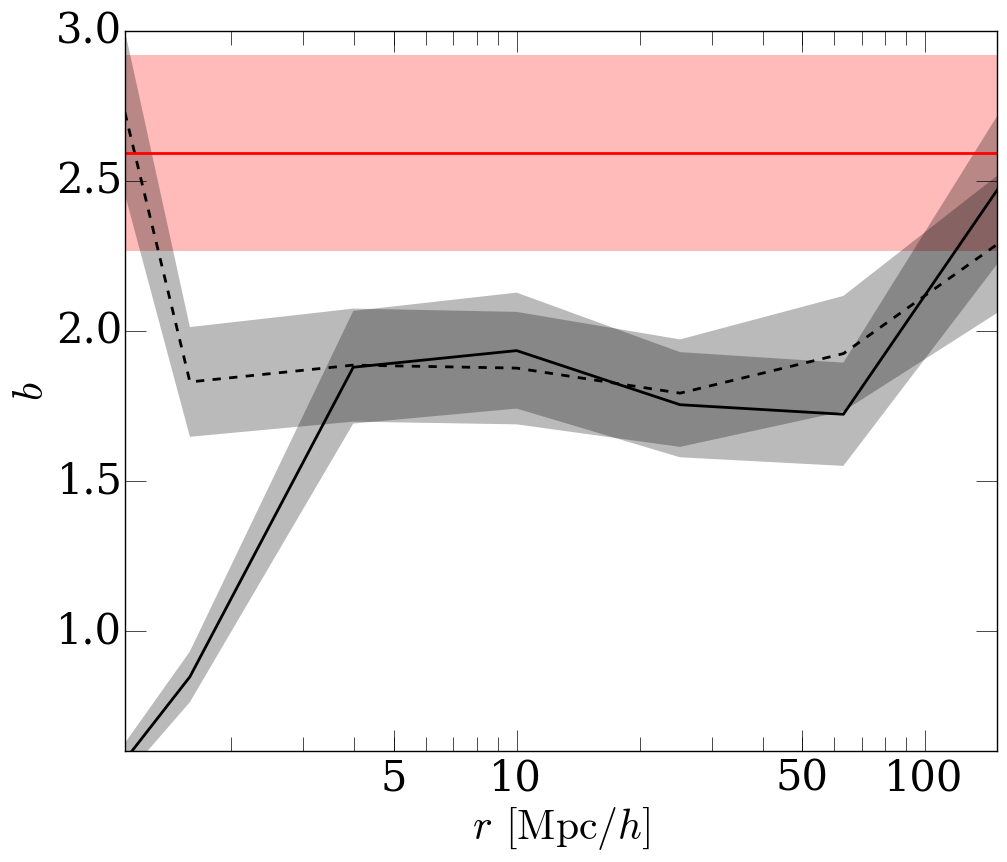}
	\includegraphics[trim=1.5cm 0. 0.2cm 0., clip, height=5.2cm]{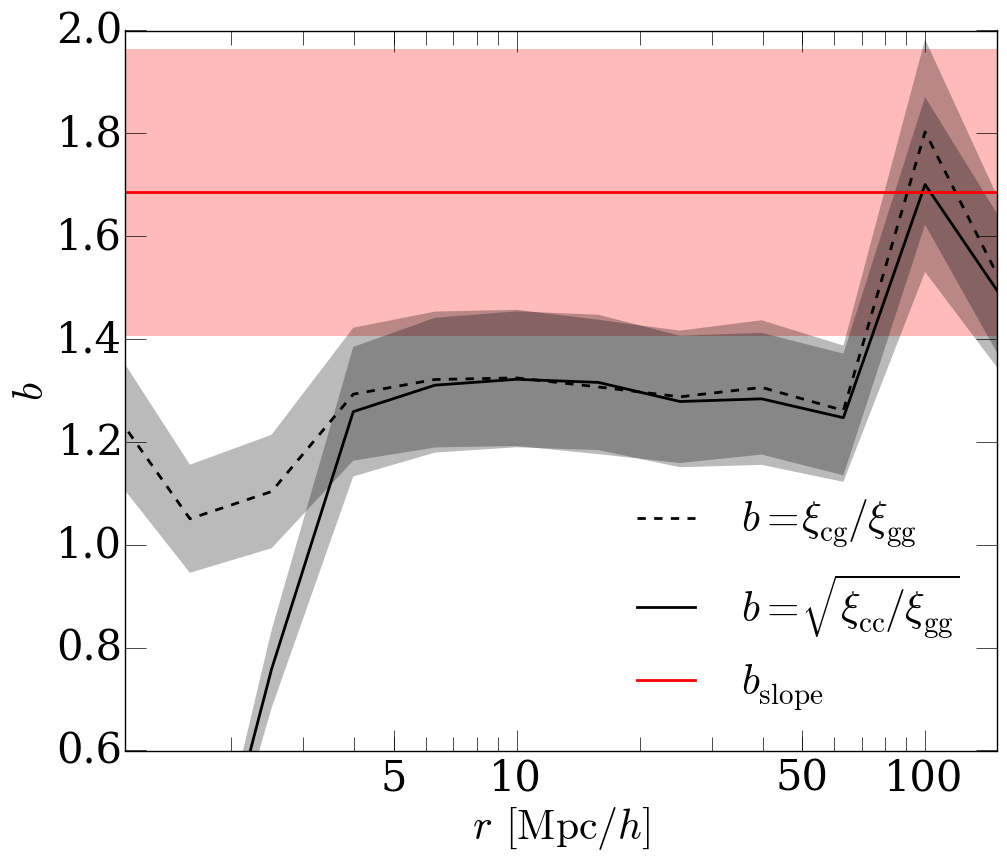}
	\includegraphics[trim=1.5cm 0. 0.2cm 0., clip, height=5.2cm]{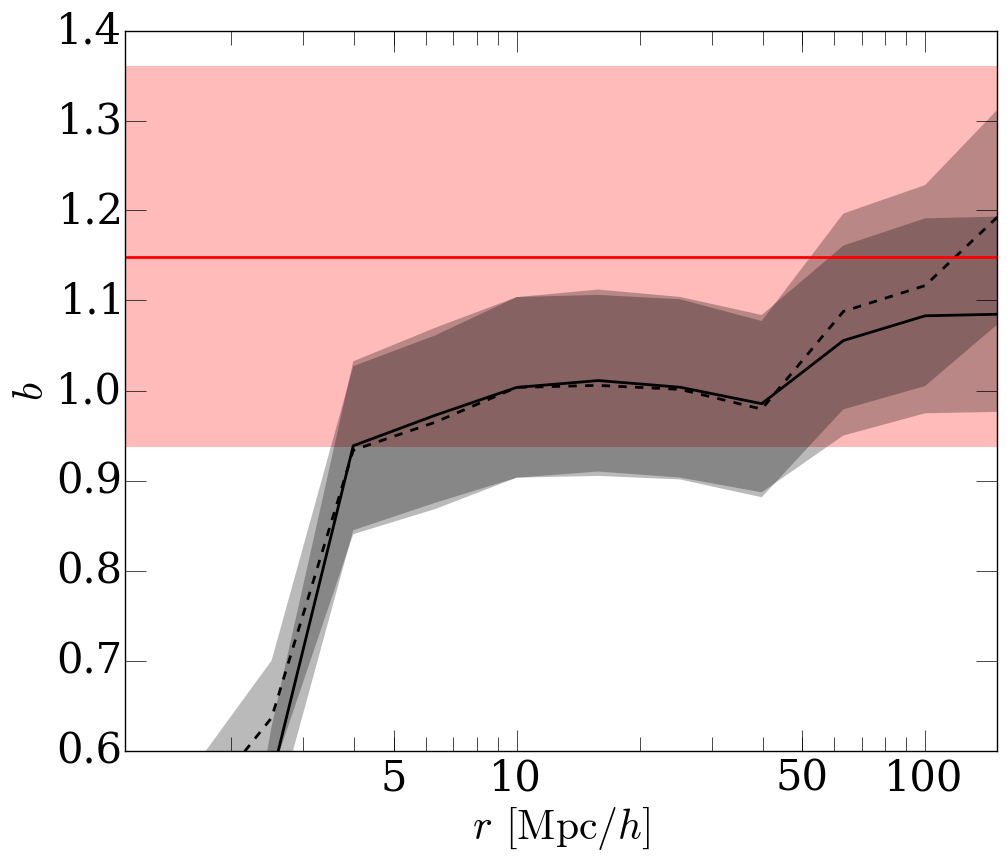}
	\caption{Comparison of the best-fit $\bs$ obtained from our largest void sample (solid red line) to the relative bias $\lrb$ between clusters and galaxies in the \magneticum\ simulation, calculated using the estimators as indicated (black dashed and dotted lines). The stellar-mass cut for the galaxy sample is varied from left to right, with the same values as in Fig.~\ref{fig:magn-bias-vs-size}.}
	\label{fig:magn-xi}
\end{figure*}

\begin{table*}
	\centering
	\caption{Best-fit values and $1\sigma$ uncertainties on the parameters of Equation~(\ref{eq:lin-fit}) for cluster-defined voids of various size and for different stellar-mass cuts in the galaxy sample from the \magneticum\ simulation.}
        \label{tab:magn-lin-fit}
        \begin{tabular}{c|cc|cc|cc} % four columns, alignment for each
		\hline
		Voids & \multicolumn{2}{c}{Galaxies ($M_{*}>1\times10^{11}h^{-1}M_{\sun}$)}  & \multicolumn{2}{c}{Galaxies ($M_{*}>5\times10^{11}h^{-1}M_{\sun}$)} &   \multicolumn{2}{c}{Galaxies ($M_{*}>1\times10^{12}h^{-1}M_{\sun}$)}  \\ 
		\hline
                \hline
                Bins in $\rv$ [$h^{-1}{\rm Mpc}$] & $b_{\rm slope}$ & $c_{\rm offset} $ & $b_{\rm slope}$ & $c_{\rm offset} $ & $b_{\rm slope}$ & $c_{\rm offset}$ \\
                \hline
                \hline
				$90 < r_{\rm v} < 110$     &$ 3.43 \pm 0.39 $ &$ -0.033 \pm 0.089$&$ 2.02 \pm 0.33  $&$  -0.013  \pm 0.095 $&$ 1.20  \pm 0.21 $&$ -0.005   \pm  0.022$ \\
				$110 < r_{\rm v} < 130 $   &$ 3.01 \pm 0.10 $ &$ -0.009 \pm 0.070$&$ 1.80 \pm 0.21  $&$  -0.003  \pm 0.064 $&$ 1.12  \pm 0.14 $&$ -0.001   \pm  0.020$ \\
				$130 < r_{\rm v} < 150 $   &$ 3.11 \pm 0.26 $ &$ -0.009 \pm 0.063$&$ 1.76 \pm 0.20  $&$  -0.005  \pm 0.063 $&$ 1.12  \pm 0.14 $&$ -0.000 \pm  0.055$ \\
				$150 < r_{\rm v} < 170 $   &$ 2.83 \pm 0.22 $ &$ -0.007 \pm 0.045$&$ 1.77 \pm 0.22  $&$  -0.003  \pm 0.063 $&$ 1.11  \pm 0.14 $&$ -0.001   \pm  0.045$ \\
                $170 < r_{\rm v} < 190 $   &$ 2.82 \pm 0.26 $ &$ -0.003 \pm 0.063$&$ 1.77 \pm 0.20  $&$  -0.001  \pm 0.061 $&$ 1.15  \pm 0.14 $&$ -0.002   \pm  0.045$ \\
                $190 < r_{\rm v} < 220 $   &$ 2.71 \pm 0.22 $ &$ -0.009 \pm 0.045$&$ 1.72 \pm 0.17  $&$  -0.004  \pm 0.060 $&$ 1.10  \pm 0.14 $&$ -0.002    \pm 0.055$ \\
                $220 < r_{\rm v} < 250 $   &$ 2.59 \pm 0.33 $ &$ -0.035 \pm 0.105$&$ 1.68 \pm 0.28  $&$  -0.017  \pm 0.101 $&$ 1.15  \pm 0.22 $&$ -0.000 \pm 0.095$ \\
                \hline
	\end{tabular}
\end{table*}

Our aim is to constrain the detailed relation between the two void density profiles. In particular, we want to check whether it is linear, similar to the relation between tracers and mass found in \cite{pollina2017}. To this end we plot the cluster-density profile $\xivc(r)$ as a function of the corresponding galaxy-density profile $\xivg(r)$ of the same cluster-voids. The results are depicted as red dots in the right panel of Fig.~\ref{fig:magn-prof}, where the error bars show the uncertainty on the mean density profiles from the left panel. The following simple linear function is used to fit those data points (black dotted line): 

\begin{equation}
\label{eq:lin-fit}
\xivc(r) = \bs\xivg(r) + \co\;,
\end{equation}
where $\bs$ and $\co$ are the only two free parameters of the fit. The linear relation between $\xivc(r)$ and $\xivg(r)$ is evident, and in concordance with the linearity between $\xivc(r)$ and the matter-density profile $\xivm(r)$ from earlier work~\citep{pollina2017}. The best-fit values for $\bs$ and $\co$, including their $1\sigma$ uncertainties can be found in Table~\ref{tab:magn-lin-fit}. $\co$ is compatible with zero within the error, while $\bs$ attains a value of about $2.7$. We expect $\bs$ to be related to the relative bias between clusters and galaxies, in analogy to Equation~(\ref{eq:relbias}).

We repeated the previous analysis for voids of different size, and confirmed the linear relation in Equation~(\ref{eq:lin-fit}) to provide a good fit in all cases. The best-fit values of $\bs$ and $\co$ are summarized in Table~\ref{tab:magn-lin-fit}. Furthermore, we explored the impact of various mass cuts in our galaxy sample. The overall clustering amplitude of galaxies is expected to depend on their stellar mass, which should be reflected in our best fit for $\bs$ as well. While our original sample contained all galaxies with stellar mass above $1\times10^{11}h^{-1} M_{\odot}$, we impose two more restrictive cuts with $M_*>5\times10^{11}h^{-1} M_{\odot}$ and $M_*>1\times10^{12}h^{-1} M_{\odot}$. Also for these cases we can confirm the linear relation of Equation~(\ref{eq:lin-fit}) to perform a good fit. The corresponding parameter constraints are reported in Table~\ref{tab:magn-lin-fit}. 

In Fig.~\ref{fig:magn-bias-vs-size} the best-fit values of $\bs$ are shown as a function of the mean effective radius of the selected void sample. The three panels correspond to the different stellar-mass cuts applied to the galaxy catalogue. We observe a clear trend of $\bs$ decreasing with void size, a similar behaviour of what has been presented in \citet{pollina2017}, albeit the different setup. In that study $\bs$ converges to a constant value for voids larger than a critical size, and this value is shown to coincide with the linear bias of the tracer with respect to the matter distribution.

In this paper, however, we are comparing the density profiles of two different tracers against each other, consequently we expect $\bs$ to converge towards the ratio of the linear bias parameters of both tracers, the linear relative bias $\lrb$. We can estimate $\lrb$ via Equation~(\ref{eq:relbias}) in two ways, both of which are plotted in Fig.~\ref{fig:magn-xi} as solid and dashed black lines with shaded error bars, respectively. On large scales both estimators agree with each other, and yield the linear relative bias between the two tracers. We compare this value with the best fit for $\bs$ obtained from the largest effective radius bin of our void sample (red solid line with shaded error bar), which is the most likely one to have converged towards $\lrb$. In the different panels of Fig.~\ref{fig:magn-xi} only the stellar-mass cut for the galaxy sample is varied, with the same values as used in Fig.~\ref{fig:magn-bias-vs-size}.

As evident from Fig.~\ref{fig:magn-xi}, the convergence of $\bs$ towards $\lrb$ is not complete in all cases. Only for the highest stellar-mass cut of $M_*>10^{12}h^{-1} M_{\odot}$ in the galaxy sample are the two values consistent with each other within the errors. At the same time, the relative bias attains the lowest value in this case, owing to the higher bias of the galaxy sample. The lower the stellar-mass cut for the galaxies, the lower becomes their bias. Therefore the relative bias between clusters and galaxies increases, which also increases the discrepancy between $\bs$ and $\lrb$. Hence, the higher the relative bias between two tracers, the larger becomes the critical void radius $\rv^+$ at which $\bs$ and $\lrb$ converge. When voids are defined in sparse tracer distributions, such as the galaxy clusters considered here, the size of $\rv^+$ may fall well beyond the range of effective void radii that can be found in the entire void sample. A similar conclusion has already been drawn in \citet{pollina2017}, where the value of $\rv^+$ was investigated for voids identified in denser tracer samples.

Nevertheless, this first test shows that the findings of \citet{pollina2017} can be indeed reproduced by measuring the relative bias with the analysis proposed in this section, which can be fully implemented with observational data.

\subsection{DES Mocks}
\label{analysis:mice}
Having confirmed a linear relationship between the densities of luminous tracers in void environments using the \magneticum\ simulation, we now want to move to more realistic data. The next step is to test our pipeline on DES mocks (\mice, see Section~\ref{mocks}), to evaluate the impact of the light cone and photometric redshift uncertainty. The latter has so far been considered as an insurmountable obstacle for the identification of three-dimensional voids, as the typical photo-z scatter of a single galaxy corresponds to line-of-sight distance errors that are comparable to the extent of most voids. This limitation lead to other innovative ideas on how to investigate the potential of voids for cosmology, which explored under-dense regions of large-scale structure in two-dimensional projections on the sky~\citep{sanchezDES2016, gruen2016}. It has been demonstrated how this approach opens up complementary ways to constrain cosmology~\citep{Barreira_etal_2015, Gruen2017, Friedrich2017, Cautun2017}. Nevertheless, as the properties of three-dimensional voids have already been extensively studied in simulations and spectroscopic surveys (see references in the introduction), it is worth testing a similar method with photometric data.

\subsubsection{Redshift uncertainty and void finding}
\label{photoz-unc}

\begin{figure}
\resizebox{\hsize}{!}{
	 \includegraphics[width=0.5\textwidth]{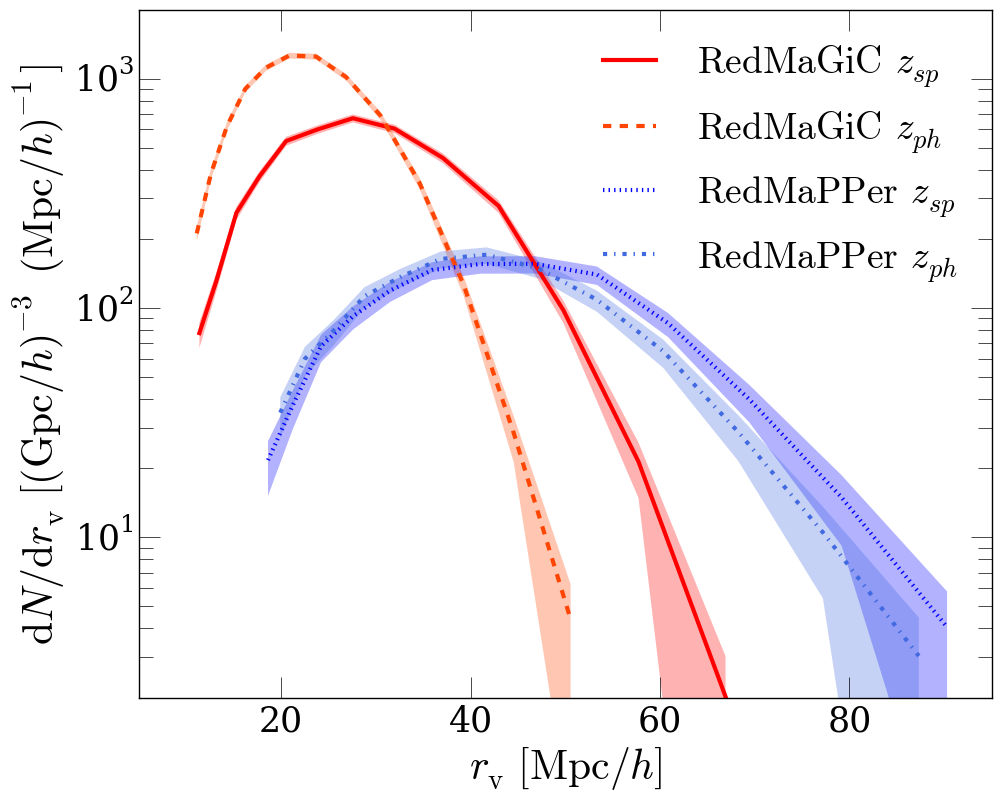}}
	\caption{The abundance of voids identified in the galaxy and cluster samples of the \mice\ mocks, as a function of their effective radius. Both photometric and spectroscopic redshifts have been used in each case, as indicated in the figure legend. The cluster-void size function is not significantly affected by photo-z uncertainty. In fact, clusters provide the most accurate photometric redshift measurements and cluster-voids are the largest voids, further reducing the relative impact of photo-z scatter on void finding.}
	\label{fig:NF-MICE2Y1}
\end{figure}

To evaluate the impact of photometric redshift uncertainty on void finding we run \vide\ on the \redmagic\ and \redmapper\ samples of the \mice\ mocks twice: once using the spectroscopic redshift (spec-z), and once the photometric (photo-z) redshift estimate of each object. The photo-z scatter inherent in the latter effects the distance estimation and causes the distribution of objects to be smeared out along the line of sight.

In Fig.~\ref{fig:NF-MICE2Y1} we present the void size function (i.e., the spatial number density of voids as a function of their effective radius) in the \mice\ mocks, extracted using \vide\ on both spectroscopic and photometric samples of galaxies and clusters. While the abundance of galaxy-voids (solid and dashed red) is heavily skewed by photo-z scatter, cluster-voids (dotted and dash-dotted blue) remain surprisingly unaffected by the choice of redshift estimate. In particular, the number of galaxy-voids with $\rvm<35h^{-1}$ Mpc is clearly overestimated when using photo-z, while the opposite is the case for larger galaxy-voids. This finding is different to what has previously been seen in \citet{sanchezDES2016}, where the largest galaxy-voids in the \redmagic\ sample were least affected by photo-z uncertainty. The disagreement is most likely a consequence of the different void finding techniques. The fact that \cite{sanchezDES2016} utilized a two-dimensional void finder on projected slices, with a line-of-sight width above the typical photo-z scatter, largely mitigates the effects of the latter. In contrast, \vide\ directly operates on three-dimensional particle distributions, and the photo-z scatter results in an unphysical line-of-sight smearing of structures that can be detected as spurious watershed ridges in the algorithm. The result is that larger voids are more likely to be segmented into multiple smaller voids.

\begin{figure*}
	\includegraphics[trim= 0. 0. 0.2cm 0., clip, height=6.9cm]{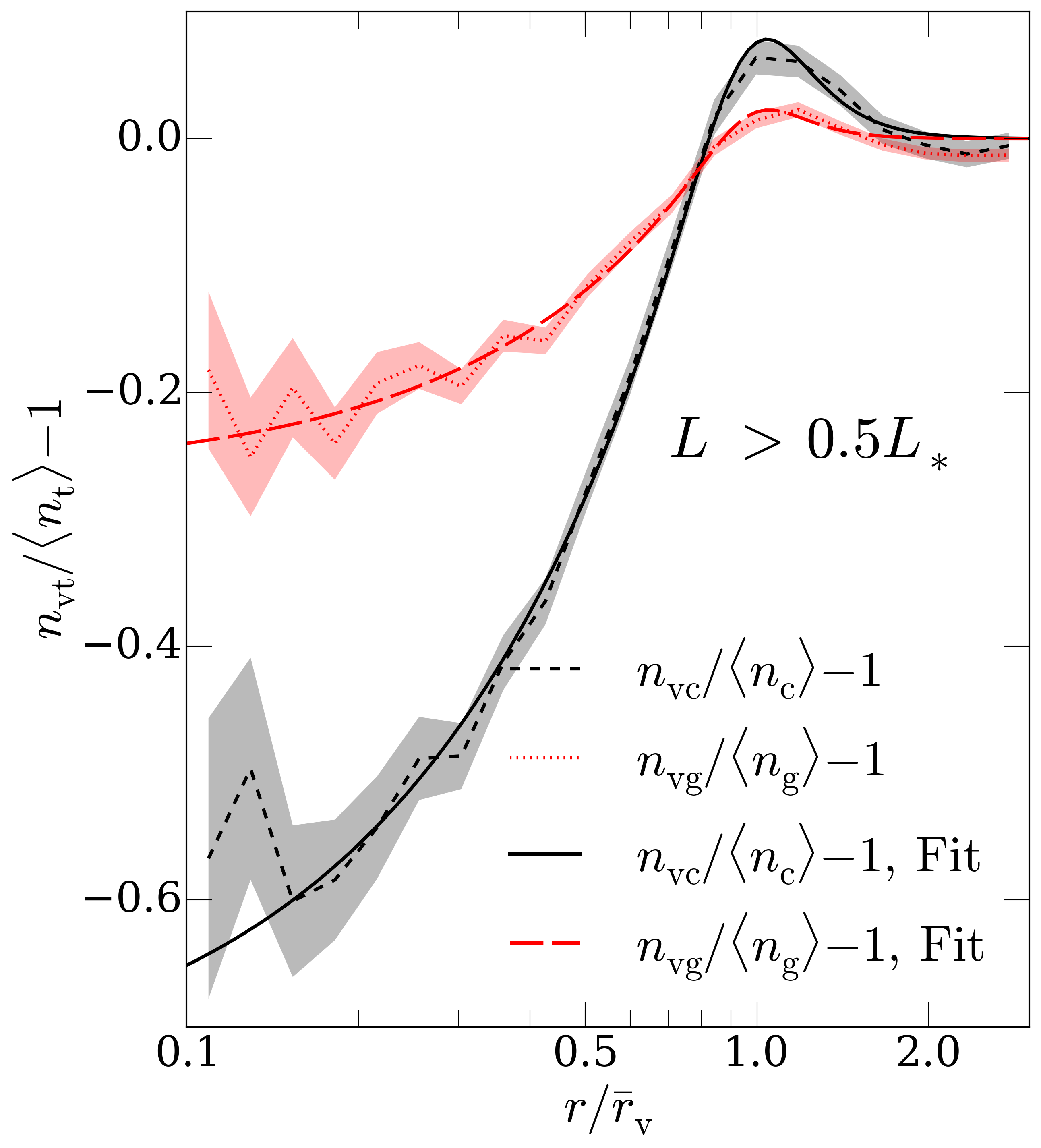}
	\includegraphics[trim = 1.9cm 0. 0.2cm 0.,clip, height=6.9cm]{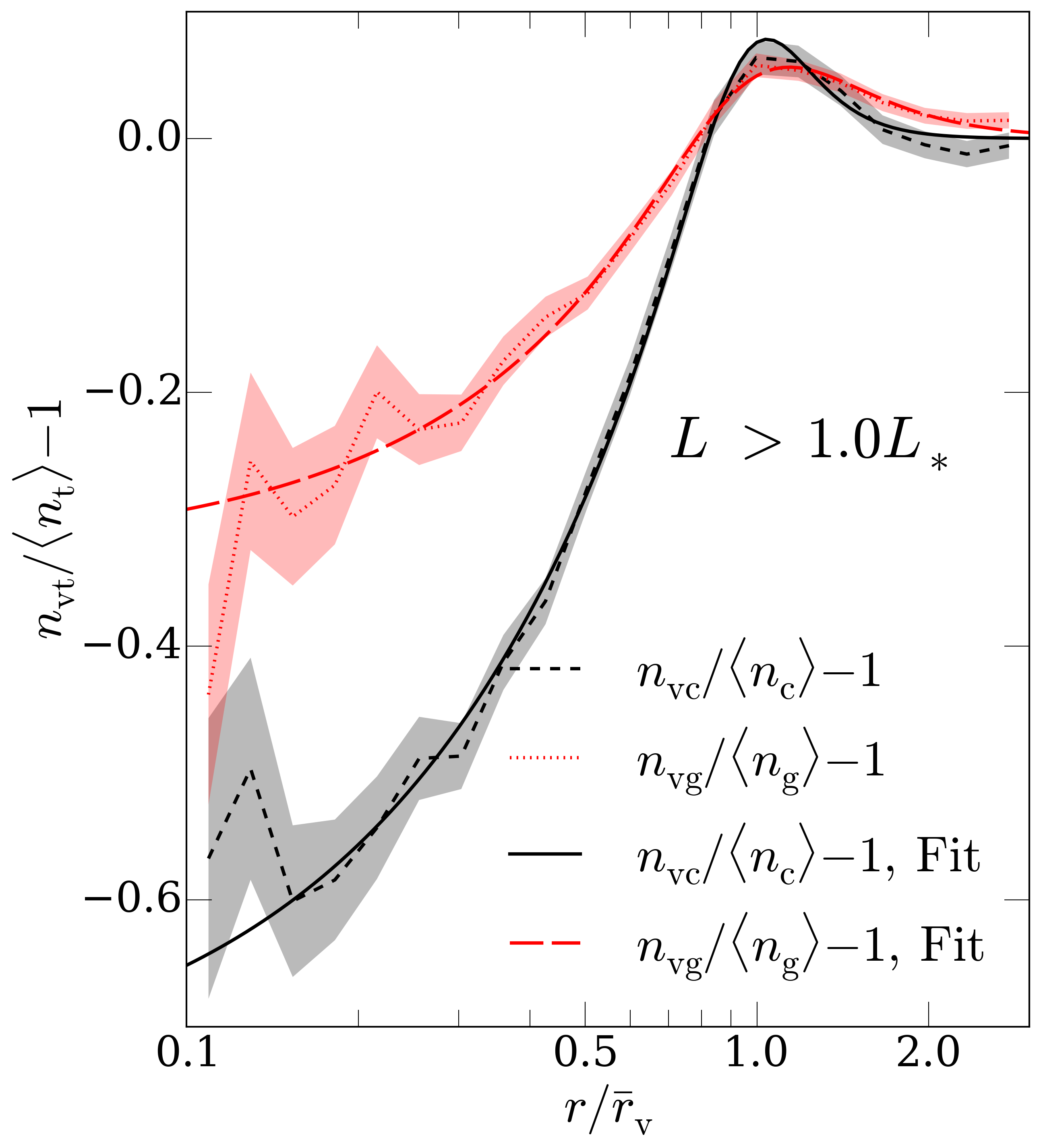}
	\includegraphics[trim = 1.9cm 0. 0.2cm 0., clip, height=6.9cm]{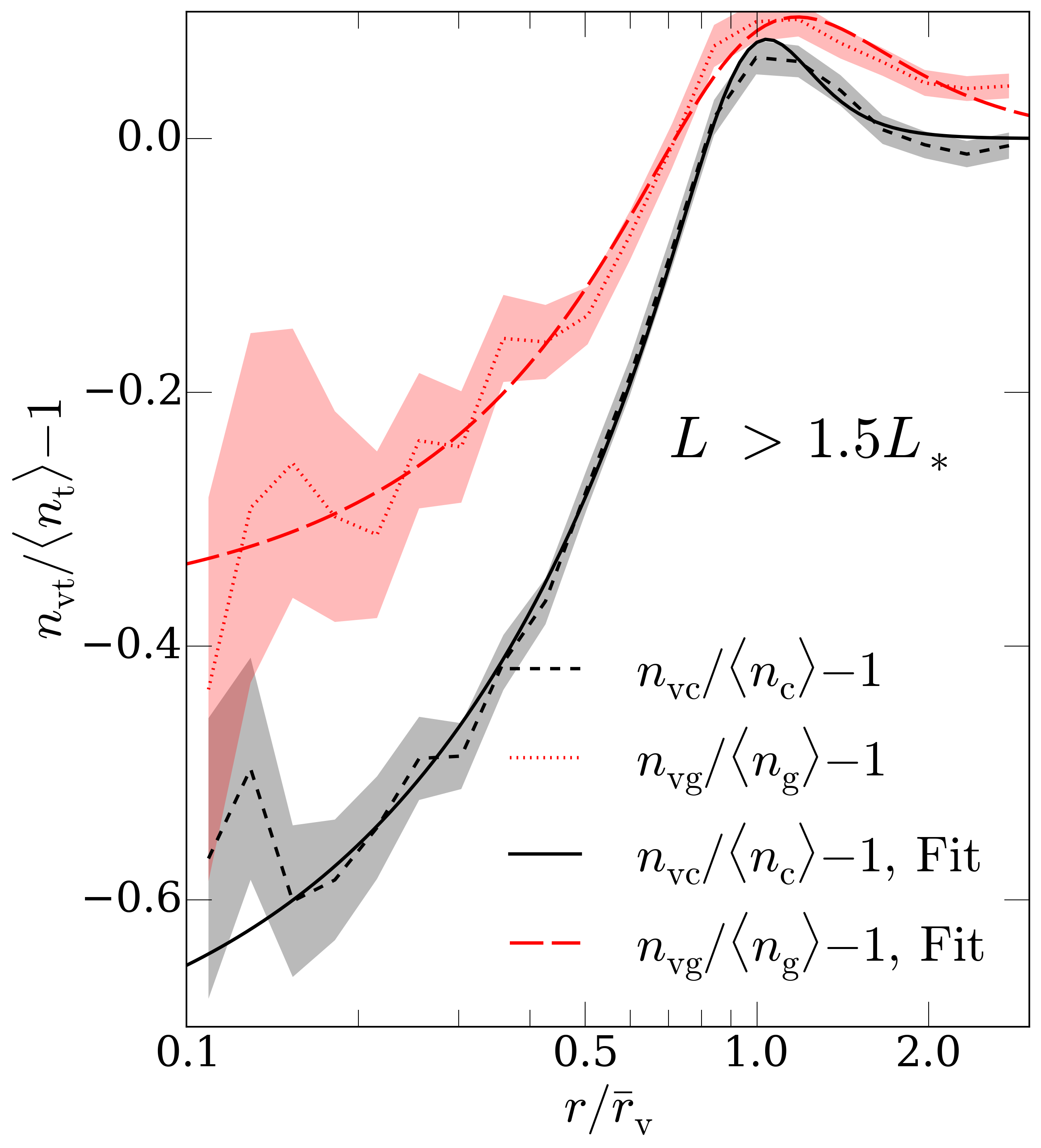}
	\caption{Tracer-density profiles (solid black for \redmapper\ clusters, dashed red for \redmagic\ galaxies) around cluster-defined voids of size $50h^{-1}$Mpc $<\rv<60h^{-1}$Mpc in the \mice\ mocks. The luminosity cut for the galaxy sample is varied from left to right, as indicated in each panel.}
	\label{fig:MICE2Y1prof}
\end{figure*}

However, this effect on void abundance is hardly detected in the cluster-void sample, thanks to the relatively accurate photometric redshift estimates in \redmapper\ clusters. The higher accuracy can be attributed to the fact that multiple member galaxies can contribute to a single cluster redshift estimate. Moreover, the sparser and more biased distribution of clusters results in larger voids overall \citep{sutter2014sparseS}, so the extent of the photo-z scatter in redshift space matters less in comparison to the void size. In order to quantify the impact of photometric redshifts on void identification in more detail, a comparison on individual voids would be needed. However, this goes beyond the scope of this paper, as we are only concerned about summary statistics here.

The robustness of the void size function from cluster-voids in the presence of photo-z scatter has promising consequences for void science with photometric surveys. For example, void number counts can be used to constrain cosmology \citep{pisani2015}, even when identified in various tracer distributions. In particular, \citet{ronconi2017} suggest a simple way to extend the prediction of void abundances to potentially observable voids: making use of Equation~(\ref{eq:voidBias}) they claim to be able to accurately forecast the void size function obtained from halos based on results from the excursion-set theory for dark matter voids. According to Fig.~\ref{fig:NF-MICE2Y1}, this method may straight-forwardly be extended to cluster-voids extracted from photometric samples, opening up to the possible exploitation of the void size function as a cosmological probe in a large variety of forthcoming surveys \citep[e.g., LSST, EUCLID, DESI, see][]{LSST,EUCLID,DESI}

\subsubsection{Density profiles and tracer bias}

\begin{figure*}
	\includegraphics[trim=0 0 0.2cm 0. clip, height=5.2cm]{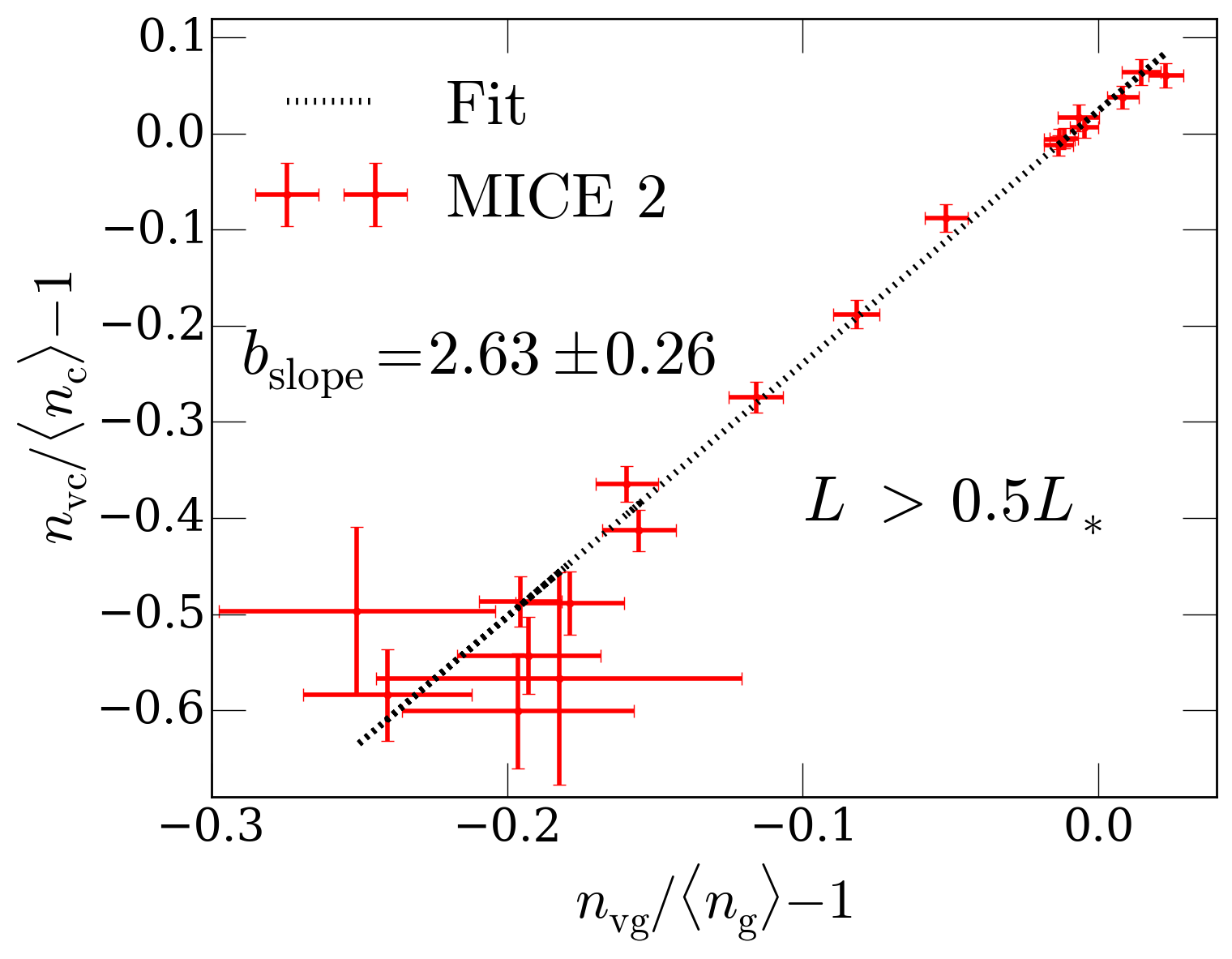}
	\includegraphics[trim= 3.3cm 0. 0.2cm 0., clip, height=5.2cm]{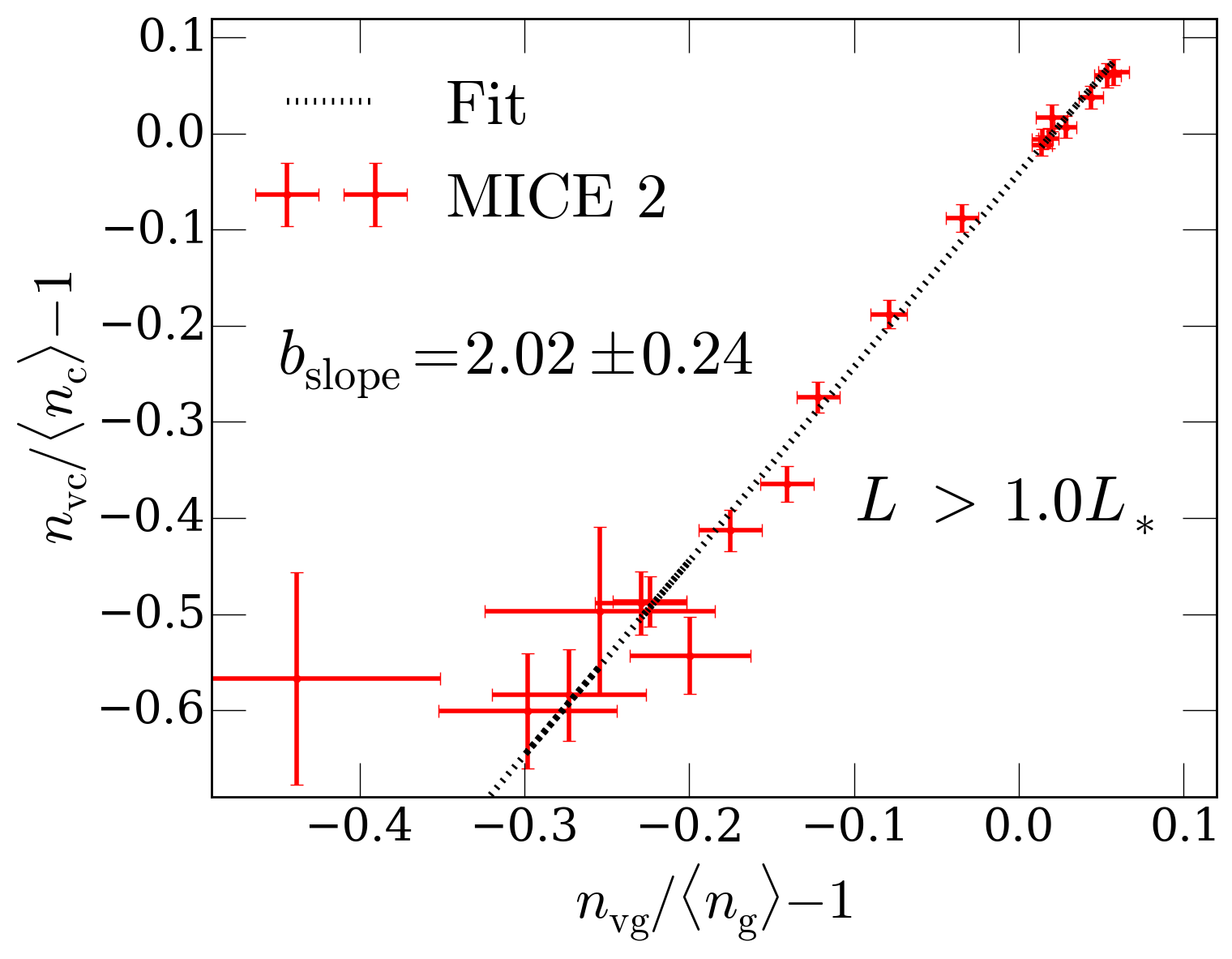}
	\includegraphics[trim =3.3cm 0. 0.2cm 0., clip, height=5.2cm]{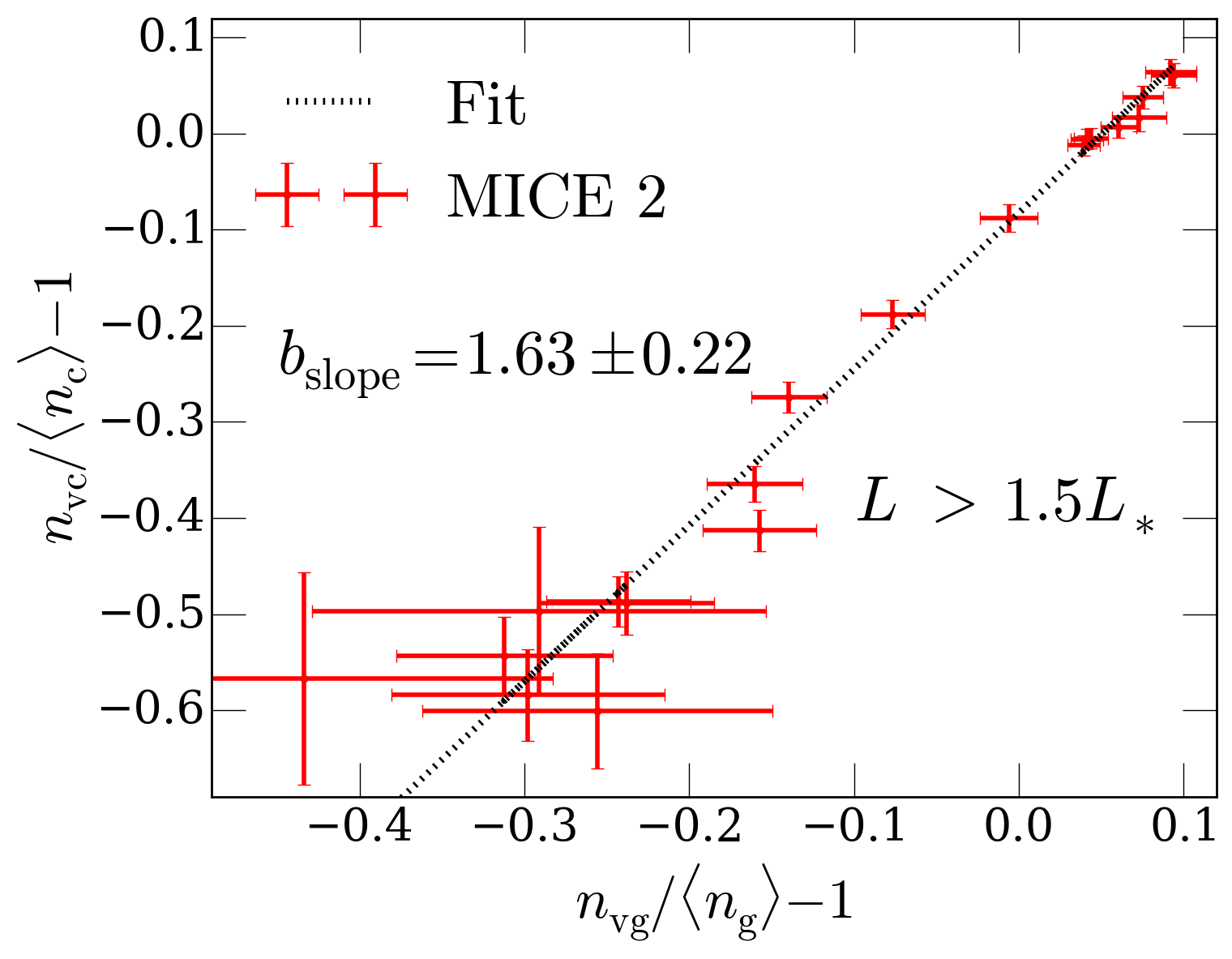}
	\caption{Cluster- and galaxy-density profiles from Fig.\ref{fig:MICE2Y1prof} plotted against each other. The dotted black line shows the best fit obtained with Equation~(\ref{eq:lin-fit}).}
	\label{fig:MICE2Y1lin}
\end{figure*}

\begin{figure*}
	\includegraphics[width=0.99\textwidth]{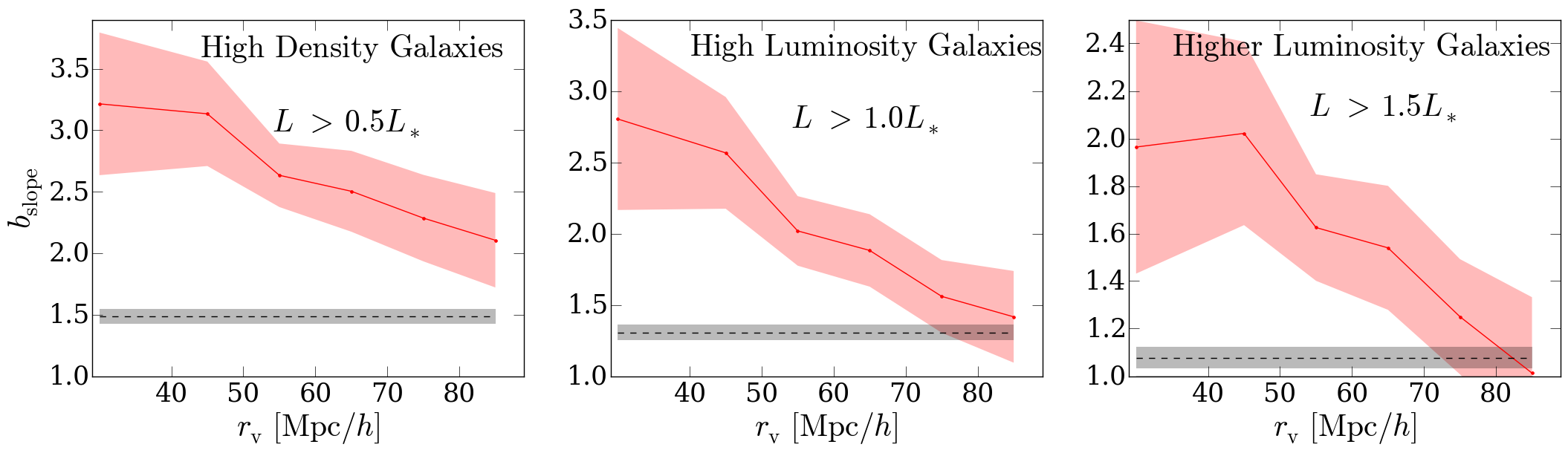}
	\caption{Best-fit values for $\bs$ (solid red) as a function of effective void radius in the \mice\ mocks. The luminosity cut for the galaxy sample is varied from left to right, as indicated in each panel. Dashed black lines show the linear relative bias between clusters and galaxies, estimated via their angular power spectra on large scales.}
	\label{fig:MICE2Y1relbias}
\end{figure*}

\begin{table*}
	\centering
	\caption{Best-fit values and $1\sigma$ uncertainties on the parameters of Equation~(\ref{eq:lin-fit}) for cluster-defined voids of various size and for different luminosity cuts in the galaxy sample from the \mice\ mocks.}
        \label{tab:mice-lin-fit}
        \begin{tabular}{c|cc|cc|cc} % four columns, alignment for each
		\hline
		Voids & \multicolumn{2}{c}{\redmagic\ ($L>0.5L_*$)}  & \multicolumn{2}{c}{\redmagic\ ($L>1.0L_*$)} &   \multicolumn{2}{c}{\redmagic\ ($L>1.5L_*$)}  \\ 
		\hline
                \hline
                Bins in $\rv$ [$h^{-1}{\rm Mpc}$] & $b_{\rm slope}$ & $c_{\rm offset} $ & $b_{\rm slope}$ & $c_{\rm offset} $ & $b_{\rm slope}$ & $c_{\rm offset}$ \\
                \hline
                \hline
				$ 20 < r_{\rm v} < 40 $   &$3.21 \pm 0.57  $ &$ -0.071\pm0.130   $&$ 2.80 \pm 0.64 $&$-0.119 \pm 0.173  $&$ 1.96  \pm 0.53 $&$ -0.084  \pm  0.158$ \\
				$ 40 < r_{\rm v} < 50 $   &$3.13 \pm 0.42  $ &$ -0.006\pm0.084   $&$ 2.56 \pm 0.39 $&$-0.066 \pm 0.089  $&$ 2.02  \pm 0.39 $&$ -0.090  \pm  0.10$ \\
				$ 50 < r_{\rm v} < 60 $   &$2.63 \pm 0.27  $ &$ 0.023 \pm0.063   $&$ 2.02 \pm 0.23 $&$-0.041 \pm 0.063  $&$ 1.62  \pm 0.22 $&$ -0.082  \pm  0.091$ \\
				$ 60 < r_{\rm v} < 70 $   &$2.50 \pm 0.33  $ &$ 0.070 \pm0.105   $&$ 1.88 \pm 0.24 $&$-0.043 \pm  0.077  $&$ 1.54  \pm 0.26 $&$ -0.111   \pm  0.183$ \\
                $ 70 < r_{\rm v} < 80 $   &$2.28 \pm 0.35  $ &$ 0.101 \pm0.126   $&$ 1.56 \pm 0.25 $&$-0.067 \pm 0.084 $&$ 1.25  \pm 0.24 $&$ -0.174  \pm  0.190$ \\
                $ 80 < r_{\rm v} < 90 $   &$2.10 \pm 0.39  $ &$ 0.162 \pm0.161   $&$ 1.41 \pm 0.32 $&$-0.127 \pm 0.128  $&$ 1.01  \pm 0.31 $&$ -0.262  \pm  0.354$ \\

                \hline
	\end{tabular}
\end{table*}

We now repeat the analysis of Section~\ref{analysis:sims} with the \mice\ mocks, using photometric redshifts for both \redmagic\ and \redmapper\ samples. The density profiles are estimated with the help of random catalogues, to account for the mask and light-cone effects. To this end, we approximate the Landy-Szalay estimator of Equation~(\ref{eq:LS}) as

\begin{equation}
\label{eq:LS2}
 \xivt(r) \simeq \langle D_{\rm v} D_{\rm t} \rangle - \langle D_{\rm v} R_{\rm t} \rangle \;,
\end{equation}
which was shown to yield accurate results on void scales~\citep{hamaus2017}. We have also compared our measurements with the more common Davis-Peebles estimator~\citep{Davis:1983aa}, which features a ratio instead of a subtraction in Equation~(\ref{eq:LS2}), and found consistent results. Fig.~\ref{fig:MICE2Y1prof} presents the corresponding tracer-density profiles for \redmapper-defined voids of size $50h^{-1}$Mpc $<\rv<60h^{-1}$Mpc. Overall we obtain smaller void sizes from this sample, as the number density of clusters here exceeds the one analysed in \magneticum. However, as \mice\ resolves galaxies of lower mass, this results in a similar relative bias between the tracers considered. As tracers, we utilize \redmapper\ clusters of richness $\lambda>5$, and three \redmagic\ samples with varying luminosity cuts. The correspondence with our earlier simulation results in the left panel of Fig.~\ref{fig:magn-prof} is striking: we observe a more pronounced cluster-density profile with a deeper core and a higher ridge (dashed black line) than each of the galaxy-density profiles (dotted red line). Yet, the shapes of all these profiles seem to match quite nicely, which means that galaxies trace voids just as the clusters do, albeit with a lower clustering amplitude. This is further confirmed by the successful interpolation of all profiles by means of the fitting function presented in Equation~(\ref{eq:prof}) (solid black and long-dashed red lines). Note that in some cases the normalization of the profiles at large distances $r$ can be slightly offset from zero. This can have various reasons, which may be related to imperfect corrections for the survey geometry, or the spread in void sizes in a given bin of $\rv$. However, we have checked that the magnitude of this effect is small enough not to impact our conclusions (i.e., $\co$ is always consistent with zero).

The correspondence between the different tracers can be seen more clearly in Fig.~\ref{fig:MICE2Y1lin}, where their void-centric density profiles are plotted against each other. A linear trend in the data is apparent, so we fit Equation~(\ref{eq:lin-fit}) and constrain its slope and offset again. We further repeat this for voids of all available sizes from our catalogue and summarize the results in Table~\ref{tab:mice-lin-fit}. The best-fit value for $\bs$ decreases when galaxies with higher luminosity cut are used. This is consistent with expectation, as they acquire a higher clustering bias, making the relative bias between clusters and galaxies decrease. In contrast, the parameter $\co$ remains consistent with zero in all cases.

The dependence of $\bs$ on void effective radius is visualized in
Fig.~\ref{fig:MICE2Y1relbias}. We observe a decreasing trend again, as before
in the \magneticum\ simulation. Towards the largest voids, $\bs$ converges to
the linear relative bias between the cluster and the galaxy samples (dashed
black line), which is estimated via the method described in
Section~\ref{bias_estimate}. However, the critical void radius $\rv^+$, where
the two relative bias measurements agree, cannot be determined from the galaxy
sample with the lowest luminosity cut. This confirms our earlier conclusion
that the convergence of $\bs$ to $\lrb$ happens at larger void radii when
$\lrb$ is higher. However, we have a clear indication that it is possible
  to measure the relative linear bias of tracers with this method when applied to the final DES dataset after 5 years of observations. We further conclude that the uncertainty inherent in photometric redshift estimates is not affecting our results from before: the linear relation of Equation~(\ref{eq:voidBias}) is still satisfied to the same degree of accuracy as in simulations, with similar constraints on its parameters.

\subsection{Data}

\begin{figure}
	\includegraphics[width=0.48\textwidth]{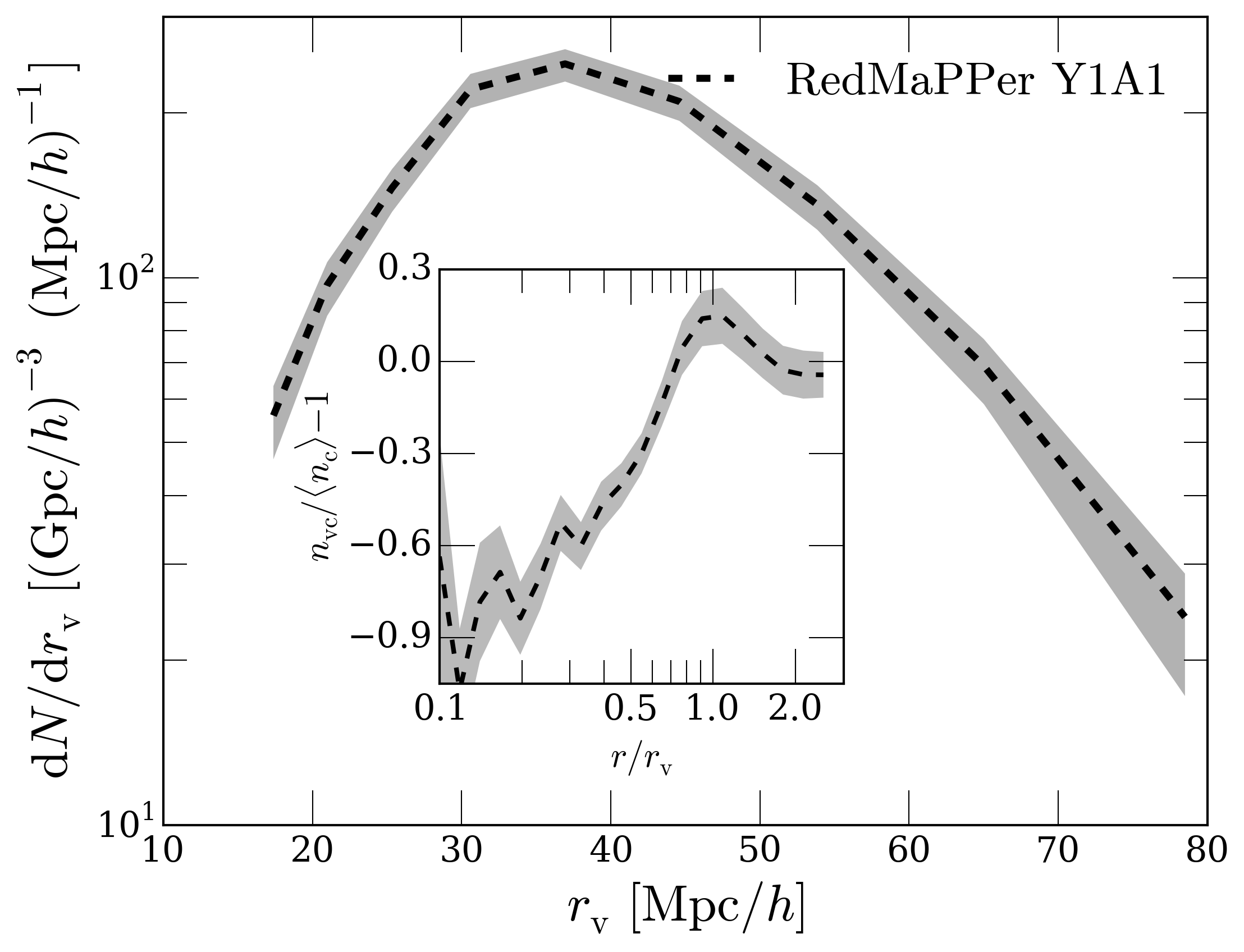}
	\caption{Abundance of voids as a function of their effective radius, identified in the distribution of \redmapper\ clusters from DES data (Y1A1). The average cluster-density profile of all voids is shown as inset.}
	\label{fig:dataNF}
\end{figure}

Having assessed the feasibility of our analysis using mocks, we are finally ready to test it on DES Y1 data and to determine whether the linear relation given by Equation~(\ref{eq:voidBias}) (applied to visible tracers) is in the sky. In this section we describe the void catalogue obtained from the data and present all related results.

\subsubsection{DES void catalogue}
\label{voidcat}

This section presents the first catalogue of three-dimensional watershed voids built with DES data. We follow our previous approach, using \redmapper\ clusters with $\lambda>5$ for void identification with \vide. Since the area observed during the first year of DES (Y1A1) operations is significantly smaller ($1321 \rm{deg}^2$) than the full octant of the \mice\ mocks, the number statistics of the data are expected to be lower. In total we find $475$ voids in the redshift range $0.2<z<0.65$ (which is the range where all \redmagic\ samples are fairly volume limited), with effective radii between $15h^{-1}$Mpc and $80h^{-1}$Mpc. Voids intersecting with the survey mask have been pruned from the final sample. The void size function is shown in Fig.~\ref{fig:dataNF}, with an inset displaying the average cluster-density profile of all voids in the sample. It is remarkably similar to that of cluster-voids in mocks shown in Fig.~\ref{fig:NF-MICE2Y1}. The small difference can be caused by the assumed mass-richness relation in the cluster mocks, which may not reproduce the real data exactly.

\begin{figure}
\resizebox{\hsize}{!}{
	\includegraphics[trim=0. 0. 2.28cm 0., clip, width=0.45\textwidth]{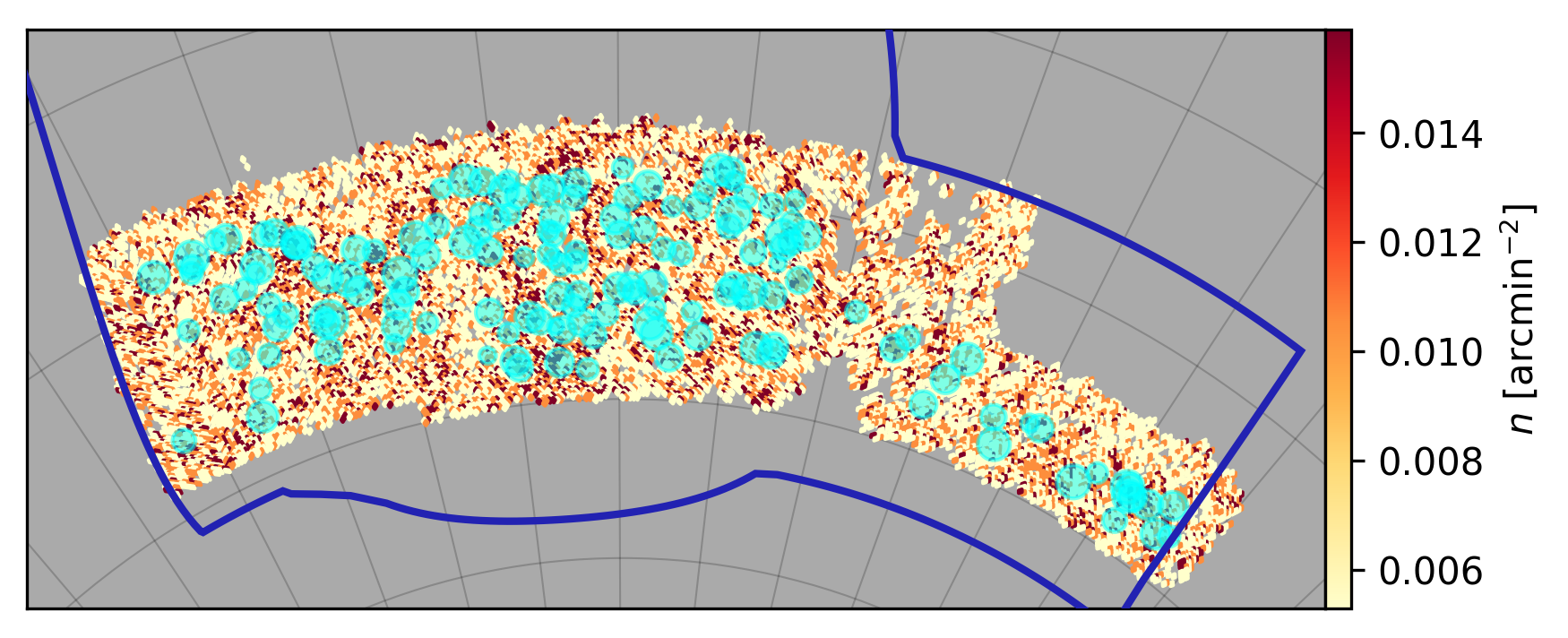}
}
	\caption{Density plot of \redmapper\ clusters and their associated void centres (cyan circles) in a redshift slice of $0.2<z<0.45$. The blue line displays the 5-year-DES footprint, voids intersecting with the survey mask are discarded.
	}
	\label{fig:map}
\end{figure}

\begin{figure}
	\includegraphics[trim=1.5cm 2.2cm 4.2cm 4.cm, clip, width=0.45\textwidth]{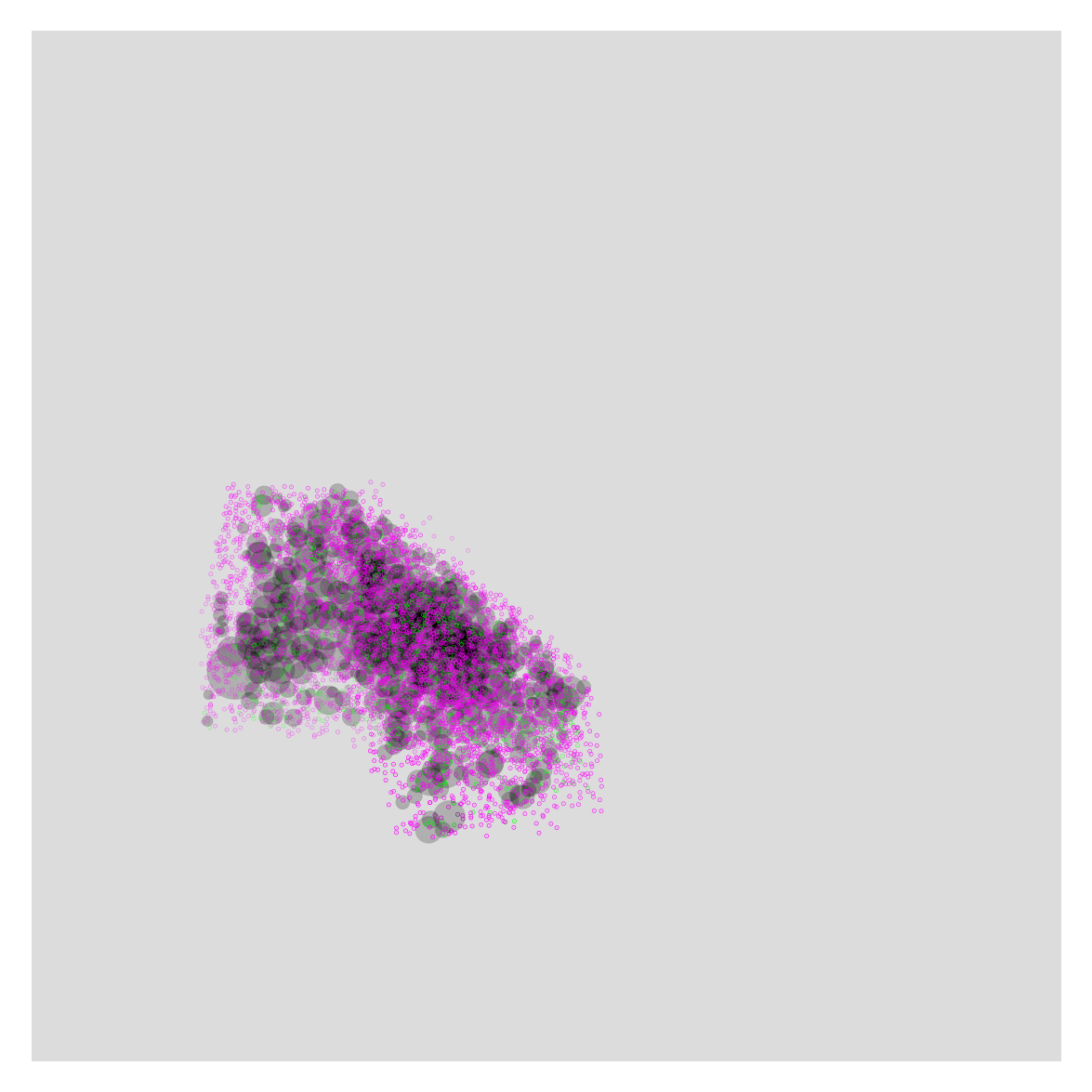}
	\caption{Three-dimensional map of the DES light cone; magenta dots show $5\%$ of all \redmapper\ clusters, green dots display $5\%$ of \redmapper\ clusters inside watershed voids and black spheres of radius $\rv$ represent the spherical volume of each void.}
	\label{fig:cone}
\end{figure}

The footprint of our void catalogue on the sky can be perceived in Fig.~\ref{fig:map}, which was made using the public code \textsc{skymapper}\footnote{\url{https://github.com/pmelchior/skymapper}}. We show the positions of void centres (cyan circles) on the density plot of clusters for a redshift slice of $0.2 < z < 0.45$. This range was chosen to allow direct comparison with Fig.~1 of \citet{Gruen2017}, where a similar map for the location of line-of-sight under-densities in the galaxy spatial distribution was presented. The blue line displays the full DES footprint at the end of its operations. Figure~\ref{fig:cone} is a three-dimensional plot of the DES light cone, where $5\%$ of all \redmapper\ clusters are shown in magenta, $5\%$ of those clusters located inside voids are highlighted in green, and black spheres of radius $\rv$ indicate the locations of void centres with a size that reflects the spherical equivalent of the watershed volume. The number of clusters was diluted for visualization purposes.

\subsubsection{Density profiles and tracer bias}

\begin{figure*}
	\includegraphics[trim= 0. 0. 0.2cm 0., clip, height=6.9cm]{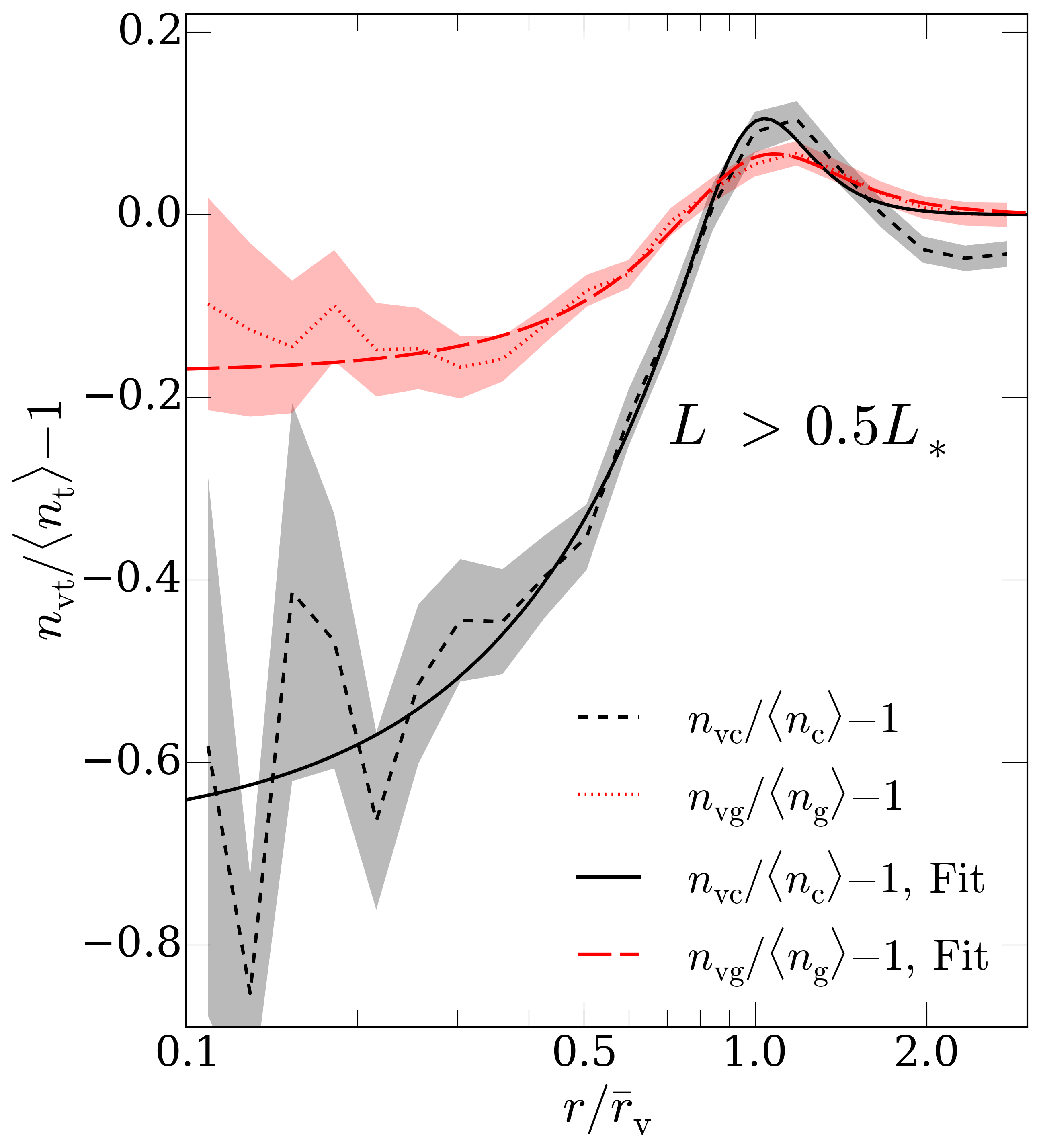}
	\includegraphics[trim = 1.9cm 0. 0.2cm 0.,clip, height=6.9cm]{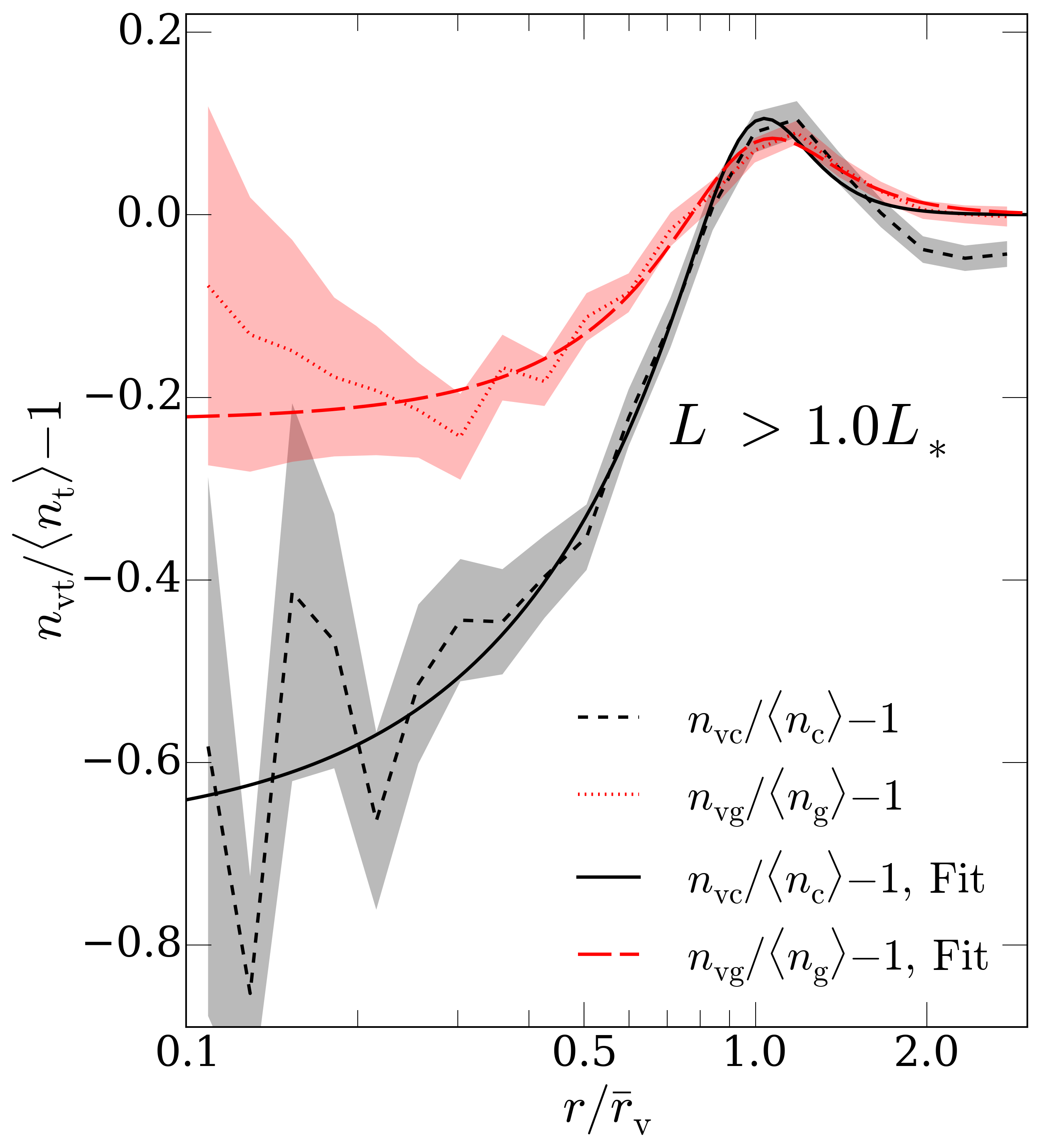}
	\includegraphics[trim = 1.9cm 0. 0.2cm 0., clip, height=6.9cm]{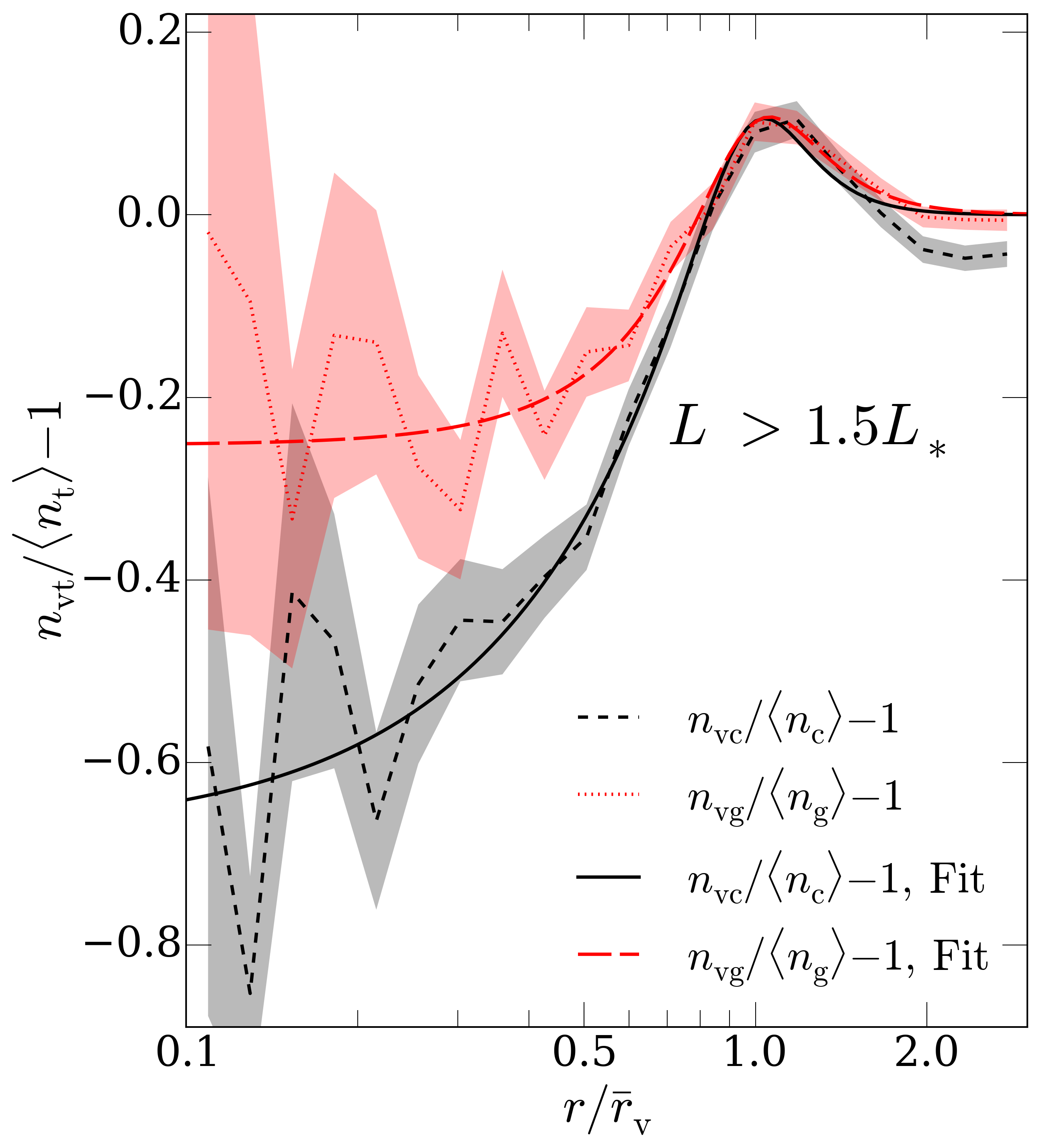}
	\caption{Tracer-density profiles (solid black for \redmapper\ clusters, dashed red for \redmagic\ galaxies) around cluster-defined voids of size $40h^{-1}$Mpc $<\rv<80h^{-1}$Mpc in the DES data. The luminosity cut for the galaxy sample is varied from left to right, as indicated in each panel.}
	\label{fig:DESY1profFull}
\end{figure*}

\begin{figure*}
	\includegraphics[trim=0 0 0.2cm 0. clip, height=5.1cm]{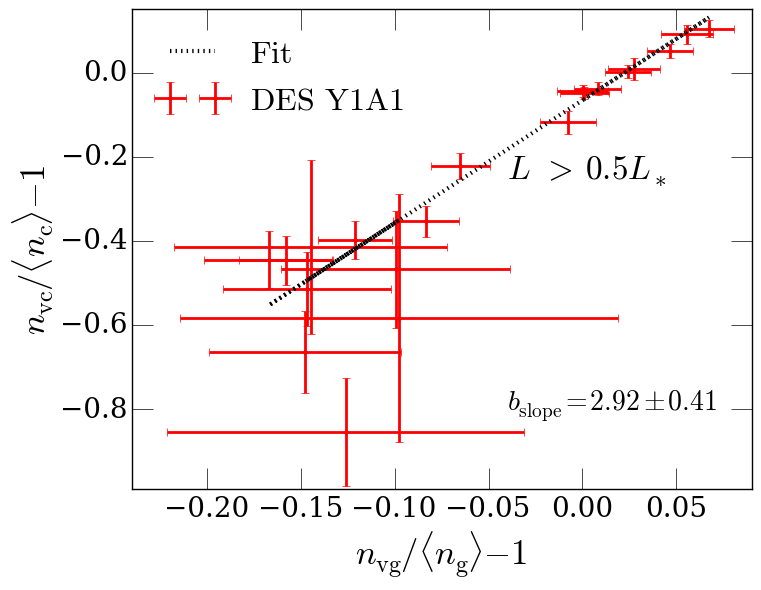}
	\includegraphics[trim= 3.3cm 0. 0.2cm 0., clip, height=5.1cm]{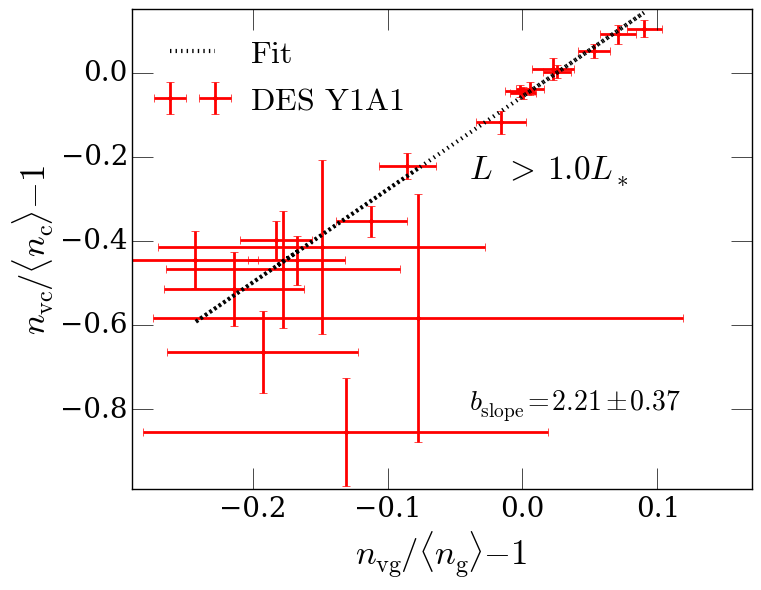}
	\includegraphics[trim =3.3cm 0. 0.2cm 0., clip, height=5.1cm]{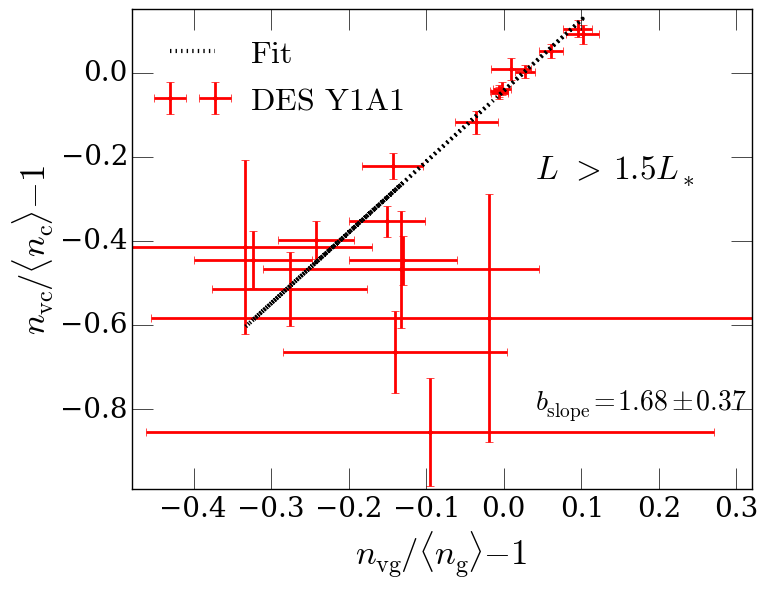}
	\caption{Cluster- and galaxy-density profiles from Fig.\ref{fig:DESY1profFull} plotted against each other. The dotted black line shows the best fit obtained with Equation~(\ref{eq:lin-fit}).}
	\label{fig:DESY1linfitFull}
\end{figure*}

\begin{figure*}
	\includegraphics[width=0.99\textwidth]{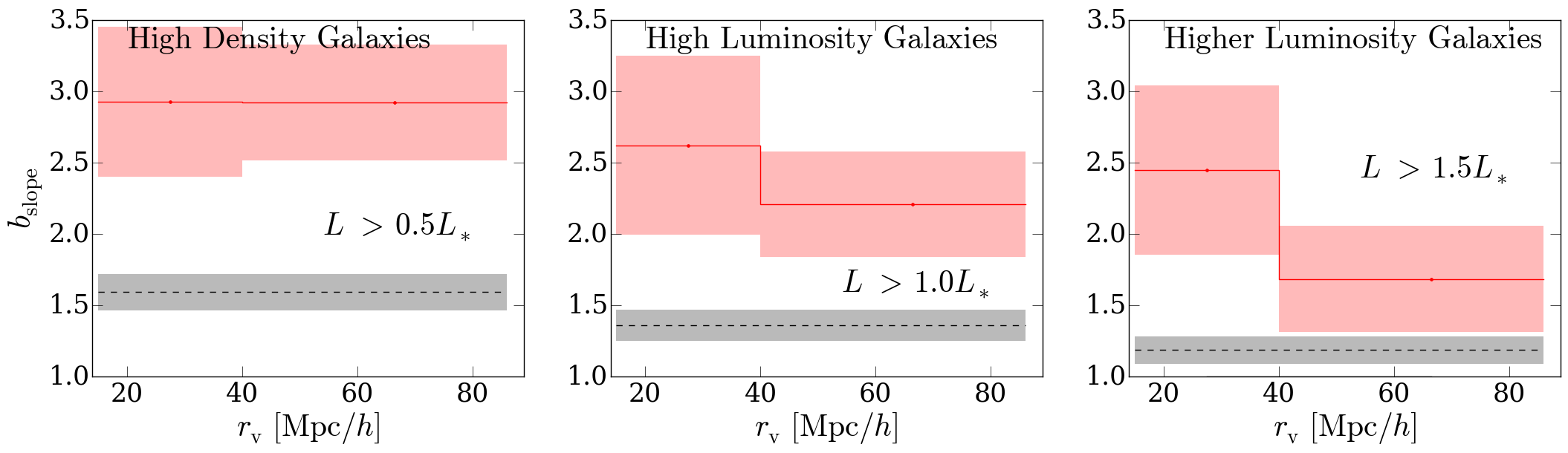}
	\caption{Best-fit values for $\bs$ (solid red) as a function of void radius in DES data. The luminosity cut for the galaxy sample varies from left to right, as indicated in each panel. Dashed black lines show the linear relative bias between clusters and galaxies, estimated via their angular power spectra on large scales.}
	\label{fig:desy1full-bias-vs-size}
\end{figure*}

With the observational void catalogue at hand, we are now in the position to apply our earlier analysis to real data. Fig.~\ref{fig:DESY1profFull} features the average tracer density profiles for cluster-voids of size $40h^{-1}$Mpc $<\rv<80h^{-1}$Mpc. As tracers, we use \redmapper\ clusters (dashed black lines) and \redmagic\ galaxies of high density, high luminosity, and higher luminosity samples (dashed red lines, from left to right). As apparent from each panel, the densities of different tracers are highly correlated in these void environments, all featuring a clear depression around the void centre, and a compensating ridge at the void edge. In particular, the similarity with the mocks in Fig.~\ref{fig:MICE2Y1prof} is striking, as is the ability of Equation~(\ref{eq:prof}) to accurately fit the data (solid black and long-dashed red lines). However, due to the smaller area it can be noted that the uncertainties in the real data are higher, especially close to the void centres, where the statistics are most affected by the sparsity of tracers.

This can also be observed in Fig.~\ref{fig:DESY1linfitFull}, where we focus on the relation between cluster- and galaxy-density profiles plotted against each other.  The linear trend in the data is apparent, although some of the data points exhibit large scatter. In all cases we find Equation~(\ref{eq:lin-fit}) to provide a satisfactory fit to the data. We find no evidence for any deviation from linearity other than due to statistical noise, which argues Equation~(\ref{eq:lin-fit}) to indeed be the simplest and most conservative model that is consistent with the data. Our earlier results based on simulations and mocks with much better statistics corroborate this result. We further confirm a decrease in the best-fit value of the slope $\bs$, caused by an increase in the bias of the galaxy samples with increasing luminosity cuts. At the same time, the offsets $\co$ remain consistent with zero. The detailed parameter constraints are reported in Table~\ref{tab:DES-lin-fit}.

Finally, we test the convergence of $\bs$ to the linear relative bias $\lrb$ of the employed tracers. Due to the relatively low number of voids in our sample, we can only afford to have two independent bins in effective radius. We choose to split the sample such that both bins roughly contain the same number of voids, with $\rv<40h^{-1}$Mpc and $\rv>40h^{-1}$Mpc. The corresponding best-fit values of $\bs$ are shown as the red dots, connected by a solid line in Fig.~\ref{fig:desy1full-bias-vs-size} (which is analogous to Fig. \ref{fig:MICE2Y1relbias} albeit with DES data). In comparison, the linear relative bias estimated via the large-scale clustering statistics of the tracers, as described in Section~\ref{bias_estimate}, is shown in dashed black. Evidently, the poor statistics in the measurement do not allow any detailed conclusions about the convergence properties of $\bs$ towards $\lrb$. However, at least for the galaxy samples of high and higher luminosity, an indication for a decrease in $\bs$ at larger $\rv$ is apparent. A more detailed investigation of this will be possible with future DES tracer catalogues of larger size. The final DES Y5 tracer catalogues will provide similar statistics as the \mice\ mocks employed above.

\section{Conclusions}
\label{Concl}
The aim of this paper was to probe the nature of tracer bias in void environments, a regime of large-scale structure that so far has little been investigated specifically for this purpose \citep[however, see][]{Neyrinck:2014aa, Yang:2017aa, Paranjape:2017aa}. In contrast, the overall tracer bias, which is typically weighted towards the most overdense structures in the Universe, has remained an active topic of research for a long time, due to its complex non-linear behaviour on intermediate and small scales \citep[e.g.,][and references therein]{Smith:2007aa, Cacciato:2012aa, Springel:2018aa, Simon:2017aa, Dvornik:2018aa}. Moreover, recent evidence for additional stochasticity beyond the Poisson expectation in the clustering properties of galaxies and clusters further complicates the common treatment of bias \citep[e.g.][]{Hamaus:2010aa, Baldauf:2013aa, Paech:2017aa, Gruen2017, Friedrich2017}. A consistent and reliable framework for the modelling of tracer bias is indispensable for the cosmological analysis of modern data sets of large-scale structure, because it establishes a connection between its observable luminous constituents and the invisible dark matter. As the latter is expected to be responsible for more than $80\%$ of the mass content in the Universe, the accuracy of cosmological constraints is often limited by the degree to which tracer bias is understood.

\begin{table*}
	\centering
	\caption{Best-fit values and $1\sigma$ uncertainties on the parameters of Equation~(\ref{eq:lin-fit}) for cluster-defined voids of various size and for different luminosity cuts in the galaxy sample from the DES data.}
        \label{tab:DES-lin-fit}
        \begin{tabular}{c|cc|cc|cc} % four columns, alignment for each
		\hline
		Voids & \multicolumn{2}{c}{\redmagic\ ($L>0.5L_*$)}  & \multicolumn{2}{c}{\redmagic\ ($L>1.0L_*$)} &   \multicolumn{2}{c}{\redmagic\ ($L>1.5L_*$)}  \\ 
		\hline
                \hline
                Bins in $\rv$ [$h^{-1}{\rm Mpc}$] & $b_{\rm slope}$ & $c_{\rm offset} $ & $b_{\rm slope}$ & $c_{\rm offset} $ & $b_{\rm slope}$ & $c_{\rm offset}$ \\
                \hline
                \hline
				$15 < r_{\rm v} < 40 $   &$ 2.92 \pm 0.53 $ &$ -0.055 \pm 0.122 $&$ 2.62 \pm 0.62 $&$ -0.038\pm 0.145 $&$ 2.45 \pm 0.59$&$ -0.027 \pm  0.138$ \\
				$40 < r_{\rm v} < 86 $   &$ 2.91 \pm 0.40 $ &$ -0.065 \pm 0.083 $&$ 2.21 \pm 0.37 $&$ -0.056\pm 0.084 $&$ 1.68 \pm 0.37$&$ -0.043 \pm  0.095$ \\
                \hline
	\end{tabular}
\end{table*}

In this work we have investigated tracer bias in void environments of the distribution of galaxy clusters, based on a complete pipeline of hydrodynamical simulations, mocks, and data from the first year of DES observations. We find a remarkably linear relationship between the void-centric density fluctuations of clusters and galaxy samples of various magnitude limits across all distance scales, suggesting tracer bias to remain linear in the two-point statistics of void environments. This confirms recent simulation results by \citet{pollina2017}, but for the first time with observational data. We show that the relative clustering amplitude between any two tracers can be expressed by a single multiplicative constant $\bs$, relating their void-tracer cross-correlation functions according to Equation~(\ref{eq:lin-fit}) with an offset consistent with zero ($\co=0$). However, the constant $\bs$ coincides with the linear relative bias $\lrb$ between those tracers only when voids above a certain critical effective radius $\rv^+$ are used in this measurement. In case of very sparse void tracers, such as the galaxy clusters used here, the value of $\rv^+$ may exceed the available range of void sizes in a given area on the sky. For smaller voids, $\bs$ increases towards lower $\rv$.

A detailed model for this behaviour can be important in cases where the absolute value of tracer bias is needed to obtain parameter constraints, which goes beyond the scope of this paper. It has been pointed out that tracer environment can be more relevant than host-halo mass to determine the bias of tracers \citep{Abbas:2007aa, Pujol:2017aa, Shi:2018aa}, and we expect the environmental constraint from voids to be important in this respect. When tracers are selected above some mass or luminosity threshold, as done here, they are typically more biased in void environments than elsewhere in the cosmic web~\citep{Yang:2017aa, Paranjape:2017aa}. Conversely, selecting the most extreme environments as tracers of the density field, such as the centres of voids, can lead to a vanishing, or even negative clustering bias~\citep{Hamaus:2014aa, clampitt2016}. Nevertheless, the fact that tracer bias can be treated linearly with a single free parameter significantly simplifies most common two-point clustering analyses of large-scale structure. For example, it implies that different tracer-density profiles around voids can be described with the same universal functional form~\citep[as provided by Equation~(\ref{eq:prof}),][]{hamaus2014, sutter2014sparseS}. The analysis we presented is arguably the best approach to test such a function, as with observational data we do not have access to the entire three-dimensional distribution of luminous and dark matter. Furthermore, the presented method can be augmented with measurements of tangential shear around voids, which provides the projected surface-mass density excess between weakly lensed source galaxies and the observer. Shape catalogues of the galaxies in DES are available, a study of the absolute tracer bias with respect to the underlying dark matter distribution in void environments is underway~(Fang et al., in prep.). 
Our conclusions are further in excellent agreement with recent analyses of weak lensing by troughs in the projected galaxy distribution~\citep{gruen2016,Gruen2017}, which can be accurately modelled using linear bias~\citep{Friedrich2017}. While those results argue for a non-vanishing stochasticity parameter to be important for the counts-in-cells statistic, this does not apply to cross-correlation functions (as employed in this paper), where stochasticity does not enter at non-zero separation.

As a side product, we have constructed the first catalogue of 3D-watershed voids that are solely based on photometric redshift measurements with a controlled photo-z uncertainty.\footnote{\cite{Granett:2008aa} have already extracted a 3D void catalogue from SDSS photometry, but analysed it in projection to study ISW imprints.} Another element of novelty in our approach is that we employ galaxy clusters, rather than single galaxies, as tracers for void finding. In fact, our tests with mocks indicate that while the accuracy of \redmagic\ redshift estimates for single galaxies is not sufficient to match void number counts from a spectroscopic survey, \redmapper\ clusters produce remarkably similar void abundances among spec-z and photo-z catalogues. The flip side of using clusters rather than galaxies as void tracers is that they can only access fewer and larger voids, due to their sparsity. Nevertheless, for our purposes this constitutes also an advantage, as the relative impact of photo-z scatter becomes even smaller for large voids. Furthermore, the high number of clusters accessible in photometric surveys opens up a promising perspective for void science in the future. In fact forthcoming surveys, such as LSST \citep{LSST} and EUCLID \cite{EUCLID}, will partially rely on photometric redshift estimates. The effort to fully exploit these kind of data in the context of void studies will thereby benefit from our analysis.

\section*{Acknowledgments}

This paper has gone through internal review by the DES collaboration. We thank Seshadri Nadathur for useful comments. GP, NH, KP, KD and JW acknowledge the support of the DFG Cluster of Excellence ``Origin and Structure of the Universe'' and the Transregio programme TR33 ``The Dark Universe''. The calculations have partially been carried out on the computing facilities of the Computational Center for Particle and Astrophysics (C2PAP) and of the Leibniz Supercomputer Center (LRZ) under the project IDs pr58we, pr83li, and pr86re. Special thanks go to LRZ for the opportunity to run the Box0 simulation within the Extreme Scale-Out Phase on the new SuperMUC Haswell extension system. We appreciate the support from the LRZ team, especially N. Hammer, when carrying out the Box0 simulation.

Funding for the DES Projects has been provided by the U.S. Department
of Energy, the U.S. National Science Foundation, the Ministry of
Science and Education of Spain, the Science and Technology Facilities
Council of the United Kingdom, the Higher Education Funding Council
for England, the National Center for Supercomputing Applications at
the University of Illinois at Urbana-Champaign, the Kavli Institute of
Cosmological Physics at the University of Chicago, the Center for
Cosmology and Astro-Particle Physics at the Ohio State University, the
Mitchell Institute for Fundamental Physics and Astronomy at Texas A\&M
University, Financiadora de Estudos e Projetos, Funda{\c c}{\~a}o
Carlos Chagas Filho de Amparo {\`a} Pesquisa do Estado do Rio de
Janeiro, Conselho Nacional de Desenvolvimento Cient{\'i}fico e
Tecnol{\'o}gico and the Minist{\'e}rio da Ci{\^e}ncia, Tecnologia e
Inova{\c c}{\~a}o, the Deutsche Forschungsgemeinschaft and the
Collaborating Institutions in the Dark Energy Survey.

The Collaborating Institutions are Argonne National Laboratory, the
University of California at Santa Cruz, the University of Cambridge,
Centro de Investigaciones Energ{\'e}ticas, Medioambientales y
Tecnol{\'o}gicas-Madrid, the University of Chicago, University College
London, the DES-Brazil Consortium, the University of Edinburgh, the
Eidgen{\"o}ssische Technische Hochschule (ETH) Z{\"u}rich, Fermi
National Accelerator Laboratory, the University of Illinois at
Urbana-Champaign, the Institut de Ci{\`e}ncies de l'Espai (IEEC/CSIC),
the Institut de F{\'i}sica d'Altes Energies, Lawrence Berkeley
National Laboratory, the Ludwig-Maximilians Universit{\"a}t
M{\"u}nchen and the associated Excellence Cluster Universe, the
University of Michigan, the National Optical Astronomy Observatory,
the University of Nottingham, The Ohio State University, the
University of Pennsylvania, the University of Portsmouth, SLAC
National Accelerator Laboratory, Stanford University, the University
of Sussex, Texas A\&M University, and the OzDES Membership Consortium.

Based in part on observations at Cerro Tololo Inter-American Observatory, National Optical Astronomy Observatory, which is operated by the Association of 
Universities for Research in Astronomy (AURA) under a cooperative agreement with the National Science Foundation.

The DES data management system is supported by the National Science Foundation under Grant Numbers AST-1138766 and AST-1536171.
The DES participants from Spanish institutions are partially supported by MINECO under grants AYA2015-71825, ESP2015-66861, FPA2015-68048, SEV-2016-0588, SEV-2016-0597, and MDM-2015-0509, 
some of which include ERDF funds from the European Union. IFAE is partially funded by the CERCA program of the Generalitat de Catalunya.
Research leading to these results has received funding from the European Research
Council under the European Union's Seventh Framework Program (FP7/2007-2013) including ERC grant agreements 240672, 291329, and 306478.
We  acknowledge support from the Australian Research Council Centre of Excellence for All-sky Astrophysics (CAASTRO), through project number CE110001020, and the Brazilian Instituto Nacional de Ci\^encia
e Tecnologia (INCT) e-Universe (CNPq grant 465376/2014-2).

This manuscript has been authored by Fermi Research Alliance, LLC under Contract No. DE-AC02-07CH11359 with the U.S. Department of Energy, Office of Science, Office of High Energy Physics. The United States Government retains and the publisher, by accepting the article for publication, acknowledges that the United States Government retains a non-exclusive, paid-up, irrevocable, world-wide license to publish or reproduce the published form of this manuscript, or allow others to do so, for United States Government purposes.

%%%%%%%%%%%%%%%%%%%%%%%%%%%%%%%%%%%%%%%%%%%%%%%%%%


\bibliographystyle{mnras}

\bibliography{bib,baldi_bibliography}

%\input{Relative_bias.bbl}

%%%%%%%%%%%%%%%%%%%%%%%%%%%%%%%%%%%%%%%%%%%%%%%%%%

%%%%%%%%%%%%%%%%% APPENDICES %%%%%%%%%%%%%%%%%%%%%
\appendix
\section{Affiliations}
\textit{
$^{1}$ Excellence Cluster Universe, Boltzmannstr.\ 2, 85748 Garching, Germany\\
$^{2}$ Universit\"ats-Sternwarte, Fakult\"at f\"ur Physik, Ludwig-Maximilians Universit\"at M\"unchen, Scheinerstr. 1, 81679 M\"unchen, Germany\\
$^{3}$ Max-Planck-Institute for Astrophysics, Karl-Schwarzschild Strasse 1, D-85748 Garching, Germany\\
$^{4}$ Max Planck Institute for Extraterrestrial Physics, Giessenbachstrasse, 85748 Garching, Germany\\
$^{5}$ Department of Physics and Astronomy, University of Pennsylvania, Philadelphia, PA 19104, USA\\
$^{6}$ Institut de F\'{\i}sica d'Altes Energies (IFAE), The Barcelona Institute of Science and Technology, Campus UAB, 08193 Bellaterra (Barcelona) Spain\\
$^{7}$ Kavli Institute for Particle Astrophysics \& Cosmology, P. O. Box 2450, Stanford University, Stanford, CA 94305, USA\\
$^{8}$ SLAC National Accelerator Laboratory, Menlo Park, CA 94025, USA\\
$^{9}$ Cerro Tololo Inter-American Observatory, National Optical Astronomy Observatory, Casilla 603, La Serena, Chile\\
$^{10}$ Fermi National Accelerator Laboratory, P. O. Box 500, Batavia, IL 60510, USA\\
$^{11}$ Institute of Cosmology \& Gravitation, University of Portsmouth, Portsmouth, PO1 3FX, UK\\
$^{12}$ Observatories of the Carnegie Institution of Washington, 813 Santa Barbara St., Pasadena, CA 91101, USA\\
$^{13}$ CNRS, UMR 7095, Institut d'Astrophysique de Paris, F-75014, Paris, France\\
$^{14}$ Sorbonne Universit\'es, UPMC Univ Paris 06, UMR 7095, Institut d'Astrophysique de Paris, F-75014, Paris, France\\
$^{15}$ Department of Physics \& Astronomy, University College London, Gower Street, London, WC1E 6BT, UK\\
$^{16}$ Laborat\'orio Interinstitucional de e-Astronomia - LIneA, Rua Gal. Jos\'e Cristino 77, Rio de Janeiro, RJ - 20921-400, Brazil\\
$^{17}$ Observat\'orio Nacional, Rua Gal. Jos\'e Cristino 77, Rio de Janeiro, RJ - 20921-400, Brazil\\
$^{18}$ Department of Astronomy, University of Illinois at Urbana-Champaign, 1002 W. Green Street, Urbana, IL 61801, USA\\
$^{19}$ National Center for Supercomputing Applications, 1205 West Clark St., Urbana, IL 61801, USA\\
$^{20}$ Centro de Investigaciones Energ\'eticas, Medioambientales y Tecnol\'ogicas (CIEMAT), Madrid, Spain\\
$^{21}$ George P. and Cynthia Woods Mitchell Institute for Fundamental Physics and Astronomy, and Department of Physics and Astronomy, Texas A\&M University, College Station, TX 77843,  USA\\
$^{22}$ Department of Physics, IIT Hyderabad, Kandi, Telangana 502285, India\\
$^{23}$ Department of Astronomy, University of Michigan, Ann Arbor, MI 48109, USA\\
$^{24}$ Department of Physics, University of Michigan, Ann Arbor, MI 48109, USA\\
$^{25}$ Institut d'Estudis Espacials de Catalunya (IEEC), 08193 Barcelona, Spain\\
$^{26}$ Institute of Space Sciences (ICE, CSIC),  Campus UAB, Carrer de Can Magrans, s/n,  08193 Barcelona, Spain\\
$^{27}$ Kavli Institute for Cosmological Physics, University of Chicago, Chicago, IL 60637, USA\\
$^{28}$ Instituto de Fisica Teorica UAM/CSIC, Universidad Autonoma de Madrid, 28049 Madrid, Spain\\
$^{29}$ Institute of Astronomy, University of Cambridge, Madingley Road, Cambridge CB3 0HA, UK\\
$^{30}$ Kavli Institute for Cosmology, University of Cambridge, Madingley Road, Cambridge CB3 0HA, UK\\
$^{31}$ Department of Physics, ETH Zurich, Wolfgang-Pauli-Strasse 16, CH-8093 Zurich, Switzerland\\
$^{32}$ Santa Cruz Institute for Particle Physics, Santa Cruz, CA 95064, USA\\
$^{33}$ Center for Cosmology and Astro-Particle Physics, The Ohio State University, Columbus, OH 43210, USA\\
$^{34}$ Department of Physics, The Ohio State University, Columbus, OH 43210, USA\\
$^{35}$ Harvard-Smithsonian Center for Astrophysics, Cambridge, MA 02138, USA\\
$^{36}$ Australian Astronomical Observatory, North Ryde, NSW 2113, Australia\\
$^{37}$ Departamento de F\'isica Matem\'atica, Instituto de F\'isica, Universidade de S\~ao Paulo, CP 66318, S\~ao Paulo, SP, 05314-970, Brazil\\
$^{38}$ Department of Astrophysical Sciences, Princeton University, Peyton Hall, Princeton, NJ 08544, USA\\
$^{39}$ Instituci\'o Catalana de Recerca i Estudis Avan\c{c}ats, E-08010 Barcelona, Spain\\
$^{40}$ Jet Propulsion Laboratory, California Institute of Technology, 4800 Oak Grove Dr., Pasadena, CA 91109, USA\\
$^{41}$ Department of Physics and Astronomy, Pevensey Building, University of Sussex, Brighton, BN1 9QH, UK\\
$^{42}$ School of Physics and Astronomy, University of Southampton,  Southampton, SO17 1BJ, UK\\
$^{43}$ Brandeis University, Physics Department, 415 South Street, Waltham MA 02453\\
$^{44}$ Instituto de F\'isica Gleb Wataghin, Universidade Estadual de Campinas, 13083-859, Campinas, SP, Brazil\\
$^{45}$ Computer Science and Mathematics Division, Oak Ridge National Laboratory, Oak Ridge, TN 37831\\
}
%%%%%%%%%%%%%%%%%%%%%%%%%%%%%%%%%%%%%%%%%%%%%%%%%%

% Don't change these lines

\label{lastpage}
\end{document}